\documentclass[AMA,STIX1COL]{WileyNJD-v2}
\pdfoutput=1
\usepackage{amsthm,amsmath}
\usepackage{graphicx}
\usepackage{bm}
\usepackage{soul}
\usepackage{diagbox}
\usepackage{url}            
\usepackage{booktabs}       
\usepackage{dirtytalk}
\usepackage{csquotes}
\usepackage{moreverb}
\usepackage{caption}
\usepackage{subcaption}
\usepackage{algpseudocode}
\usepackage{algorithm}
\usepackage{mathtools}

\newcommand{\party}[2]{\frac{\partial #1}{\partial #2}} %
\renewcommand{\d}{\mathrm{d}}
\renewcommand{\b}[1]{\boldsymbol{#1}}

\articletype{RESEARCH ARTICLE}%

\received{}
\revised{}
\accepted{}
        
\begin{document}

\title{A verified and validated moving domain CFD solver with applications to cardiovascular flows}

\author[1]{Henrik A. Kjeldsberg}
\author[1]{Joakim Sundnes}
\author[1]{Kristian Valen-Sendstad}

\authormark{KJELDSBERG \textsc{et al}}

\address{
    \orgdiv{Department of Computational Physiology},
    \orgname{Simula Research Laboratory}, 
    \orgaddress{\state{Oslo}, \country{Norway}}}

\corres{
    *Kristian Valen-Sendstad, Department of
    Computational Physiology, Simula
    Research Laboratory, Oslo, Norway. 
    \email{kvs@simula.no}}
    
\fundingInfo{
This research was supported by the ERACoSysMed PARIS project, and the SimCardioTest project, and has received funding from   
the European Union's Horizon 2020 research and innovation programme under Grant Agreements No.~643271, and No.~101016496, respectively.}

\abstract[Abstract]{
Computational fluid dynamics (CFD) in combination with patient-specific medical images has been used to correlate flow phenotypes with disease initiation, progression and outcome, in search of a prospective clinical tool.
A large number of CFD software packages are available, but are typically based on rigid domains and low-order finite volume methods, and are often implemented in massive low-level C++ libraries. 
Furthermore, only a handful of solvers have been appropriately verified and validated for their intended use.
Our goal was to develop, verify and validate an open-source CFD solver for moving domains, with applications to cardiovascular flows.
The solver is an extension of the CFD solver Oasis, which is based on the finite element method (FEM) and implemented using the FEniCS open source framework. 
The new solver, named OasisMove, extends Oasis by expressing the Navier-Stokes equations in the arbitrary Lagrangian-Eulerian formulation, which is suitable for handling moving domains. 
For code verification we used the method of manufactured solutions for a moving 2D vortex problem, and for validation we compared our results against existing high-resolution simulations and laboratory experiments for two moving domain problems of varying complexity. 
Verification results showed that the $L^2$ error followed the theoretical convergence rates. The temporal accuracy was second order, while the spatial accuracy was second and third order using 
$\mathbb{P}_1/\mathbb{P}_1$ and $\mathbb{P}_2/\mathbb{P}_1$ finite elements, respectively. 
Validation results showed good agreement with existing benchmark results, by reproducing lift and drag coefficients with less than 1\% error, and demonstrating the solver's ability to capture vortex patterns in transitional and turbulent-like flow regimes. 
In conclusion, we have shown that OasisMove is an open-source, accurate and reliable solver for cardiovascular flows in moving domains. 
}
\keywords{Computational fluid dynamics, arbitrary Lagrangian-Eulerian formulation, moving domain, verification, validation, left ventricle, hemodynamics}
\maketitle

\footnotetext{\textbf{Abbreviations:} CVD, cardiovascular diseases; CFD, computational fluid dynamics; ALE, arbitrary Lagrangian-Eulerian; MMS, method of manufactured solutions; N-S, Navier-Stokes; LV, left ventricle; ED, end-diastolic; ES, end-systolic; IPCS, incremental pressure correction scheme; VMTK, Vascular Modeling Toolkit; V\&V, verification and validation; CT, computed tomography; MRI, magnetic resonance imaging; CFL, Courant–Friedrichs–Lewy}

\section{Author note}
This is the pre-peer reviewed version of the following article \cite{kjeldsberg2023verified}, which has been published in final form at \url{https://doi.org/10.1002/cnm.3703}. This article may be used for non-commercial purposes in accordance with Wiley Terms and Conditions for Use of Self-Archived Versions.

\section{Introduction}
\label{introduction}
Cardiovascular diseases (CVD) are the leading cause of global mortality and morbidity, being responsible for 33.5\% of all deaths worldwide in 2019, and is a major economic burden to society~\cite{roth2020global}. 
Many CVDs are tightly linked to the hemodynamic forces exerted on the cardiovascular system, which drive pathological remodeling of the vasculature~\cite{bassiouny1994hemodynamic,meng2007complex,nixon2010critical}.
However, medical imaging techniques such as computed tomography (CT) and magnetic resonance imaging (MRI) cannot directly capture the hemodynamic forces, and direct reconstruction of the flow field from medical images is often inaccurate~\cite{potters2014measuring,prevrhal2011ct}. 
To overcome these limitations, a combination of medical imaging tools and computational fluid dynamics (CFD) has been employed to seek a correlation between abnormal cardiovascular stresses and the initiation, progression, and outcome of diseases~\cite{schirmer2007prediction,sforza2009hemodynamics}.
This strategy has been successful in domains where the deformations are relatively small~\cite{valen2014high}, for instance in smaller blood vessels, which can therefore be approximated as rigid domains. 
However, CFD has more recently been applied in moving domains such as larger arteries and the left heart~\cite{jia2019image,sanatkhani2021subject}, where the assumptions of rigid walls may lead to limited physiological insight. 
This group of applications gives rise to a new class of cardiovascular CFD problems, where flows are studied in moving domains. 
A particular feature of these cardiovascular flow problems is that the deformation of the vasculature is often adequately captured by medical imaging, and it is therefore highly relevant to study CFD problems in domains with a known, prescribed motion. 

A large number of CFD solvers are available, with variable software design, including solvers that have support for moving domains through techniques such as the immersed boundary method~\cite{peskin2002immersed} or the arbitrary Lagrangian-Eulerian~\cite{hirt1974arbitrary} (ALE) method, making the choice of solver non-trivial.
On one hand, commercial solvers, such as Ansys Fluent~\cite{matsson2021introduction} or COMSOL~\cite{pryor2009multiphysics}, provide low-level implementations, let the user handle simulations through a convenient graphical interface, and provide quick and robust solution methods. 
However, the user remains limited by operating a black-box software, and default settings trade accuracy for speed\cite{ansys2011ansys}, which may lead to overly simplified numerical solutions.
On the other hand, open-source solvers such as OpenFOAM~\cite{jasak2007openfoam} or LifeV~\cite{bertagna2017lifev} consist of flexible low-level implementations where the user has full control of the solver details. 
However, such tools often require the user to be familiar with the implemented numerical method, resulting in a steeper learning curve, and designing flow problems can become a cumbersome and time-consuming task.
Furthermore, although only performed for a small selection of problems, comparative studies show considerable variability in the
results~\cite{jeong2014comparison,dhunny2016numerical,welahettige2016comparison,valen2018real}.
Finally, although extensively verified for 2D solutions, CFD software rarely undergo verification and validation (V\&V) for their intended context of use~\cite{us2016reporting,v2018assessing}.
Hence, certain solvers may potentially not satisfy the FDA-recognized standards for verification, validation, and uncertainty assessment~\cite{asme2022vandv}.

To address user-friendliness, availability, reproducibility, and credibility, we consider the verified and validated open-source rigid domain CFD solver Oasis~\cite{mortensen2015oasis}.
Oasis is based on the open-source finite element framework FEniCS~\cite{logg2012automated}, with a high-level Python interface.
The flow problems in Oasis are implemented as Python scripts, allowing the user to control every aspect of the solver and problem parameters in only a few lines of code.
The solver has previously been validated against various spectral element solvers for both aneurysmal~\cite{steinman2013variability} and stenotic~\cite{khan2019direct} flow.
In addition, Oasis has been validated \textit{in vitro} FDA nozzle
benchmark, where jet breakdown location, including potential transition to
turbulence, was precisely captured for Reynolds numbers ranging from 500 to 6500~\cite{bergersen2019fda}.
Furthermore, Oasis has shown to be robust for a broad set of problems, including flow in the internal carotid artery~\cite{valen2013high,valen2014mind,khan2017non}, and the respiratory system~\cite{farghadan2019topological}. 
Thus, the aim of this study was to develop, verify, and validate an open-source CFD solver that allows for prescribed boundary motion and mesh movement. 
For this purpose we have modified and extended Oasis' functionality, to support moving domains and solving the Navier-Stokes equation in the ALE formulation.  

The paper is organized as follows. 
In Section \ref{method} we present the ALE formulation of the Navier-Stokes equations and outline the solution method as well as the methods and test cases for V\&V. 
Section \ref{results} presents qualitative and quantitative results of the V\&V, while sections \ref{discussion} and \ref{conclusion} discusses and concludes on the implications of the work.

\section{Method}
\label{method}
\subsection{The Navier-Stokes equations in the arbitrary Lagrangian-Eulerian formulation}
We consider the incompressible Navier-Stokes (N-S) equations in a domain $\Omega \subset \mathbb{R}^d$, $d=2,3$ with a sufficiently smooth boundary $\partial\Omega = \Gamma$:
\begin{alignat}{2}
  \party{\b u}{t}\bigg|_{\b X} &= -\frac{1}{\rho} \nabla p + \nu \nabla^2 \b u + \b f &\qquad \text{in } \Omega, \label{eq:ns1}\\[0.8em]
  \nabla \cdot \b u &= 0  &\qquad \text{in } \Omega, \label{eq:ns2}
\end{alignat}
where $\b u =\b u( \b x, t)$ is the velocity vector, $p=p(\b x, t)$ is the pressure, $\rho$ is the fluid density, $\nu$ is the kinematic viscosity, and $\b f$ represents any body forces. 
Here, $|_{\b X}$ denotes the material time derivative following a fluid particle $\b X$, and the spatial derivatives are taken with respect to the spatial (Eulerian) coordinates $\b x$. 

Classical fluid mechanics in rigid domains is typically based on the Eulerian formulation in \eqref{eq:ns1}-\eqref{eq:ns2}, where the computational mesh is fixed and the fluid moves relative to the grid.
We will formulate and solve the equations in moving domains, and for this purpose we employ the ALE formulation ~\cite{hirt1974arbitrary,souli2000ale}. 
The ALE formulation combines the classical Eulerian formulation of fluid flow with the Lagrangian approach commonly used in solid mechanics, where the coordinates move with the deformation. 
To obtain the ALE formulation of the N-S equations, we need to replace the material time derivative of \eqref{eq:ns1} with a time derivative taken with respect to a fixed point of the mesh located in an arbitrary reference frame.
As introduced above, we have the Eulerian coordinates $\b x$ describing a point in the Euclidean space, and the material coordinates $\b X$ describing a fixed fluid particle.
We now introduce the mesh coordinate $\b \chi$ describing a fixed point of the mesh, such that $\b x(\b X, t)$ and $\b x(\b \chi, t)$ describe the position of a fluid element and mesh point in the Eulerian coordinates, respectively. 
The fluid and mesh velocity can then be defined as:
\begin{equation}
    \b u = \party{\b x}{t}\bigg|_{\b X}, \quad   \b w = \party{\b x}{t}\bigg|_{\b \chi},  \label{eq:u-w-vels}   
\end{equation}
and the material time derivative for a fluid velocity field $\b u = \b u(\b x, t)$ can be expressed as:
\begin{equation}
    \party{\b u(\b x(\b X, t), t)}{t}\bigg|_{\b X} = \party{\b u}{t}\bigg|_{\b x} + (\b u\cdot \nabla)\b u. \label{eq:u-conv}
\end{equation}
Furthermore, the time derivative relative to the arbitrary reference frame $\b \chi$ becomes:
\begin{equation}
    \party{\b u(\b x(\b \chi, t), t)}{t}\bigg|_{\b \chi} = \party{\b u}{t}\bigg|_{\b x} + (\b w \cdot \nabla)\b u, \label{eq:w-conv}
\end{equation}
where the spatial derivatives are taken with respect to the Eulerian coordinates $\b x$. 
Inserting \eqref{eq:u-conv} and \eqref{eq:w-conv} into \eqref{eq:ns1}, and adding suitable boundary conditions, results in the following ALE formulation of the N-S equations:
\begin{align}
      \party{\b u}{t} \bigg|_{\b \chi} + (\b u -\b w) \cdot \nabla \b u &= -\frac{1}{\rho} \nabla p + \nu \nabla^2 \b u + \b f
      &\qquad \text{in } \Omega(t), 
      \label{eq:ns1-ale}\\
      \nabla \cdot \b u &= 0  &\qquad \text{in } \Omega(t), \label{eq:ns2-ale}\\
       \b u &= \b g &\qquad \text{on } \Gamma^{\text{W}}(t), \label{eq:ns3-ale}\\ 
       \b u &= \b f &\qquad \text{on } \Gamma^{\text{IN}}(t), \label{eq:ns4-ale}\\ 
       \b \sigma (\b u, p) \b n &= -p_0 \b n &\qquad \text{on } \Gamma^{\text{OUT}}(t). \label{eq:ns5-ale}
\end{align}
Here, $\Gamma^{\text{W}}, \Gamma^{\text{IN}}$ and $ \Gamma^{\text{OUT}}$ define the wall, inlet and outlet boundaries, $\b g = \b g(\b x, t)$ and $\b f = \b f(\b x,t)$  are expressions describing the mesh and fluid velocity, $\b \sigma$ is the Cauchy stress tensor, and $p_0 = 0$. 
Note that without loss of generality we will assume that $\b x(\b \chi, t=0) = \b \chi$, and also use the notation $\Omega(t=0) =\Omega_0$ for the computational domain at the initial time. 

Note that \eqref{eq:ns1-ale} and \eqref{eq:ns2-ale} are valid for all interior and boundary points within $\Omega(t)$.
Typically, the motion is only prescribed at the boundary  $\Gamma(t)$, and therefore we need to compute a suitable mesh velocity and deformation for the interior of $\Omega(t)$.  
A common approach is to apply a harmonic extension of $\b w$ from the boundary to the interior domain, by solving the Laplace equation~\cite{masud2007adaptive}:
\begin{align}
     \nabla \cdot (\gamma \nabla \b w) &= 0 \qquad \text{in } \Omega(t), \label{eq:move}\\ 
    \b w &= \b g \qquad \text{on }  \Gamma(t).
\end{align}
Here, $\b g= \b g(\b x, t)$ describes the mesh velocity at the boundary and is prescribed as a Dirichlet boundary condition. 
The function $\gamma$ introduces mesh stiffening, causing the mesh to deform non-uniformly, and can be tuned to obtain a suitable mesh deformation. 
In this study we have set $\gamma$ to depend on the finite element cell volume, as previously proposed~\cite{jasak2006automatic}.

\subsection{Discretization of the Navier-Stokes equations in the ALE formulation}
To solve the N-S equations we will be using an incremental pressure correction scheme (IPCS), which splits \eqref{eq:ns1-ale} and \eqref{eq:ns2-ale} into smaller problems, and solves for the velocity and pressure in a segregated manner. 
Temporal discretization is performed by a finite difference scheme, while spatial discretization is done with finite elements. 
Following Simo and Armero~\cite{armero1992new}, a general second-order numerical scheme can be written as:
\begin{align}
  \frac{u_k^{\star}- u_k^{n-1}}{\Delta t} + B_k^{n-1/2} &=  - \nabla_k p^{\star} + \nu {\nabla}^2\tilde{u}_k+ f_k^{n-1/2} \qquad  \text{for } k = 1,\dots,d, \label{eq:fs1}\\[0.8em]
  {\nabla}^2 \varphi &= -\frac{1}{\Delta t}\nabla \cdot \b u^{\star},\label{eq:fs2} \\[0.8em]
  \frac{u_k^n - u_k^{\star}}{\Delta t} &=- {\nabla}_k \varphi \qquad \qquad \qquad \qquad \qquad \,\text{for } k = 1,\dots,d,\label{eq:fs3}
\end{align}
where $u_k^n = u(x_k,t^n)$ and $\nabla_k$ are the $k$-th velocity and gradient component, $u^{\star}$ is the tentative velocity, $t^n = n\Delta t$ denotes time, and $d$ is the number of spatial dimensions. 
The non-linear convective term $B_k^{n-1/2}$ is linearized with an Adams-Bashforth projection:
\begin{align}
  B_k^{n-1/2} = (\b u_k^{n-1/2} - \b w_k^n) \cdot \nabla u^{n-1/2} \approx \left(\frac{3}{2}\b u^{n-1} - \frac{1}{2}\b u^{n-2} - \b w^n \right) \cdot \nabla u^{n-1/2}, \label{eq:ab}
\end{align}
where $u^{n-1/2}$ is defined using semi-implicit interpolation to keep the second order accuracy in time and to avoid strict time step restrictions: 
\begin{align}
  u^{n-1/2} \approx \frac{u^{\star} + u^{n-1}}{2}. \label{eq:cn}
\end{align}
Furthermore, $\varphi = p^{n-1/2} - p^{\star}$ is the pressure correction, $p^{\star}$ is the previous tentative pressure, and we have made the variable change $p = \frac{1}{\rho}\tilde{p}$, where $\tilde{p}$ is the physical pressure in \eqref{eq:ns1-ale}. 

The ALE-based N-S solver is implemented as an extension of the CFD solver Oasis~\cite{mortensen2015oasis}, and will in the following be referred to as \textit{OasisMove}. The solver is available as open-source software on GitHub~\cite{kjeldsberg2022oasismove}.
The complete solution procedure is outlined in Algorithm \ref{alg:oasismove}. First the solver is initiated by specifying the boundary conditions and the maximum number of inner iterations – \texttt{maxiter}.
\footnote{Preliminary results indicate that using \texttt{maxiter}=2 reduced the error sufficiently. We performed preliminary analysis on the number of inner iterations using the vortex problem described in Section \ref{sec:vortex}. We compared the absolute $L^2$ error in velocity, and results showed that two inner iterations were sufficient to reduce the error below $10^{-5}$. Absolute errors below $10^{-6}$ could be achieved by performing up to 10 iterations, but at the cost of additional computational resources.}
Each time step is then initiated by solving \eqref{eq:move} for the mesh velocity $\b w$, and moving the mesh accordingly. 
Then the solver proceeds by solving \eqref{eq:fs1} for the tentative velocity, and \eqref{eq:fs2} for the pressure correction iteratively, for the selected number of inner iterations. 
Finally, \eqref{eq:fs3} is solved for the velocity correction.

\begin{algorithm}
\caption{Numerical scheme for solving the N-S equations in moving domains.} 
  \label{alg:oasismove}
\begin{algorithmic}
\State set boundary conditions
\State $t=0$
\State $\texttt{maxiter}=2$
\While{$t < T$}
    \State $t = t + \Delta t$
    \State solve \eqref{eq:move} for $ w_k^n$ for $k=0,\dots,d$
    \State update mesh coordinates
    \For{(\texttt{iter = 0; iter < maxiter; iter++})}
        \State $\varphi = p^{\star} = p^{n - 1/2}$
        \State solve \eqref{eq:fs1} for $u_k^{\star}$ for $k=0,\dots, d$
        \State solve \eqref{eq:fs2} for $p^{n-1/2}$
        \State $\varphi = p^{n-1/2} - \varphi$
   \EndFor
   \State solve \eqref{eq:fs3} for $u_k^n$ for $k=0,\dots,d$
   \State update to next time step
\EndWhile
\end{algorithmic}
\end{algorithm}

\subsection{Verification and validation}
The methods to assess the accuracy and reliability of a CFD solver, or other computational models, are typically referred to as \emph{verification} and \emph{validation}~\cite{oberkampf2002verification}.
We follow the definitions provided by the \textit{Society for Computer Simulation}~\cite{schlesinger1979terminology}, which define verification as the assessment of a numerical model's ability to represent a conceptual model with a specified accuracy, while validation is the assessment of the model's applicability and accuracy for a given application. 
A common and somewhat simplistic definition is that verification ensures that the equations are solved correctly, while validation ensures that the correct equations are solved.

A common approach for code verification is the method of manufactured solutions~\cite{roache2002code} (MMS), which is based on specifying a solution \textit{a priori}, and then deriving analytical source terms or boundary terms to make the solution match the governing equations.
Numerical errors can then be computed directly from the solutions of the solver and the specified analytical solution, and the order of accuracy of the solver can be assessed by studying how the error is reduced as either temporal or spatial resolution is increased. 
The implementation is considered verified if the known theoretical convergence rates of the applied numerical schemes are achieved. 
Model validation, on the other hand, requires that the obtained solutions are compared with reference data from high-accuracy numerical solutions as well as physical experiments, to demonstrate that the solution captures the important behavior of the relevant problems. 
We have chosen one benchmark case for verification and two cases for validation of OasisMove, all involving moving domains. 
The verification is performed using MMS, while for validation we perform qualitative and quantitative comparisons against numerical and experimental reference data.

\subsection{Description of verification and validation problems}
\subsubsection{2D Verification – Vortex problem with oscillating boundary}
\label{sec:vortex}
We considered a two-dimensional vortex problem inspired by the Taylor Green vortex as described by Fehn et al.~\cite{hesthaven2007nodal}
The problem solves the N-S equations in the absence of body forces, with the following manufactured solution:
\begin{align}
    \b u(\b x,t) &= (-\sin (2\pi  x_2), \sin(2\pi x_1))e^{-4\nu \pi^2 t},\label{eq:verify-u1} \\[0.8em]
    p(\b x,t) &= -\cos(2\pi x_1)\cos(2\pi x_2)e^{-8\nu \pi^2 t}, \label{eq:verify-u2}
\end{align}
where $\b x = (x_1, x_2)$ are the Euclidean coordinates, and $\nu$ = 0.025 m$^2$/s. 
On each of the four sides, we applied a Dirichlet boundary condition, determined by the analytical solution. 
Furthermore, the mesh velocity $\b w$ was defined by the following displacement field:
\begin{align}
    \b x(\b \chi,t) &= \b \chi + A \sin \left(\frac{2 \pi t}{T_G}\right) \left(\sin\left(2\pi \frac{\chi_2 + L / 2}{L}\right),\sin\left(2\pi \frac{\chi_1 + L / 2}{L}\right)\right), \label{eq:verify-w1} \\[0.8em]
    \b w (\b \chi, t) &= \party{\b x}{t}, \label{eq:verify-w2}
\end{align}
where $\b \chi = (\chi_1, \chi_2)$ are the ALE coordinates, $A=0.08$ is the amplitude, $T_G=4T$ is the period length of the mesh motion, where $T$ is the length of one cycle, and $L=1$.
The problem was solved on a unit square mesh, initially defined by $\Omega_0 = [-L/2, L/2]^2$, and was deformed according to \eqref{eq:verify-w1}. 
The simulations were run over the time interval $t \in (0,T]$, with $T=1.0$ and $T=2.0$ for the spatial and temporal convergence test, respectively.
For the spatial convergence tests, simulations were performed with piecewise linear ($\mathbb{P}_1$), and piecewise quadratic ($\mathbb{P}_2$) elements for the velocity, and $\mathbb{P}_1$ elements for the pressure, and we set the time step $\Delta t = 10^{-3}$ to practically eliminate temporal integration errors.
For the temporal convergence tests, we ensured that spatial discretization errors were negligible by using higher order elements; piecewise quartic ($\mathbb{P}_4$) and piecewise cubic ($\mathbb{P}_3$) elements for velocity and pressure, respectively.

\begin{figure}
     \centering
     \begin{subfigure}[b]{0.40\textwidth}
         \includegraphics[width=\linewidth]{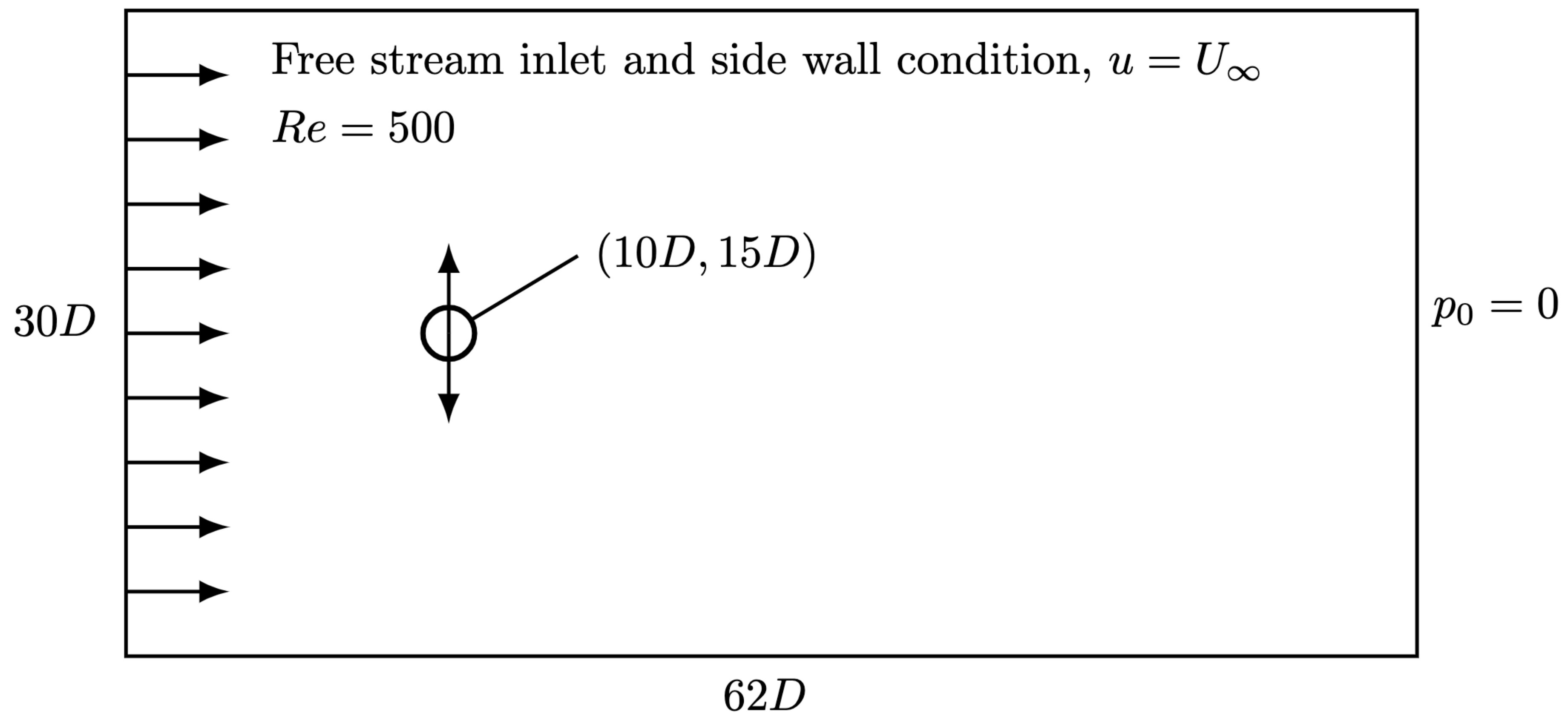}
         \caption{}
     \end{subfigure}
     \begin{subfigure}[b]{0.38\textwidth}
         \includegraphics[width=\linewidth]{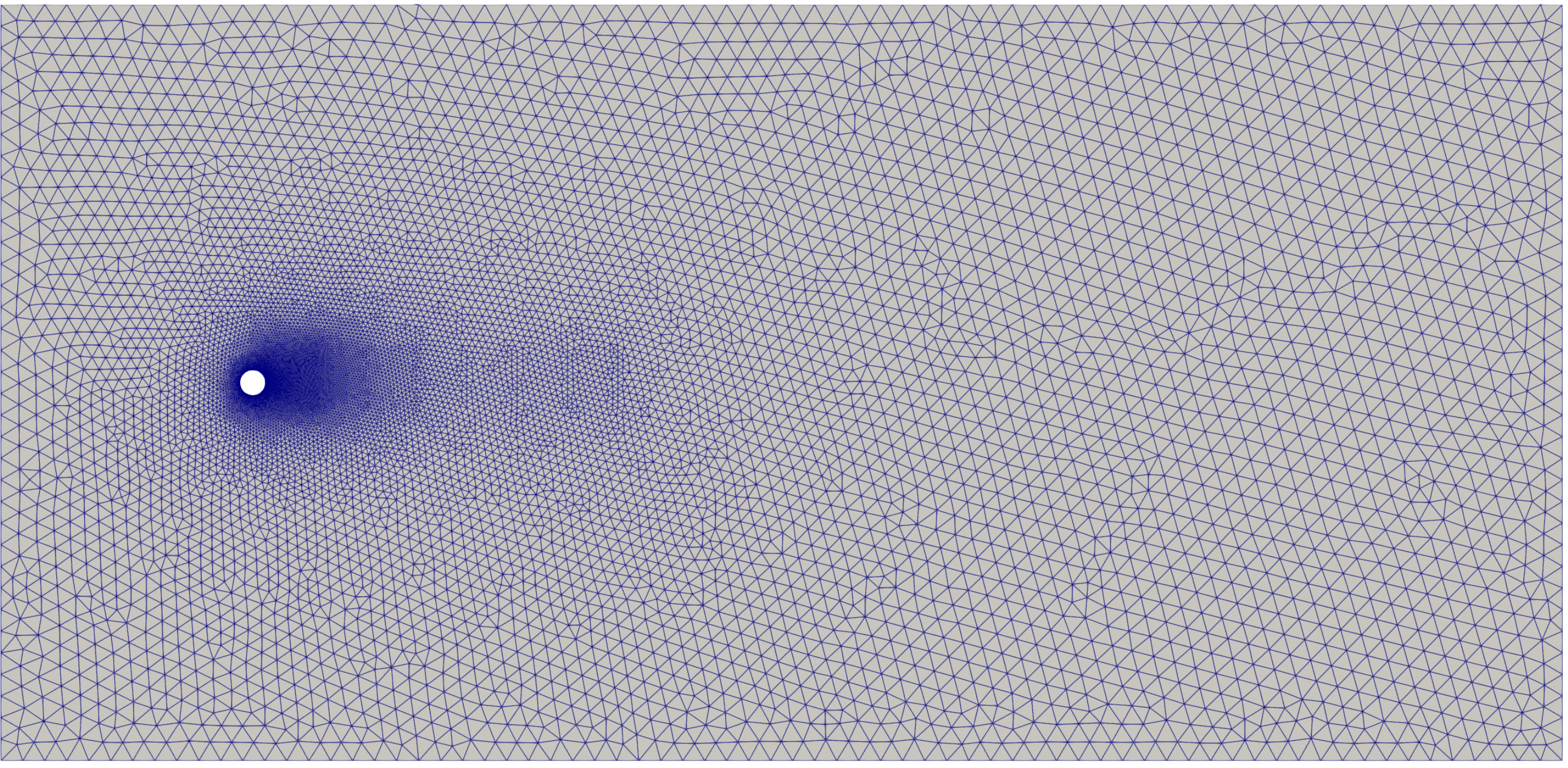}
         \caption{}
     \end{subfigure} 
     \begin{subfigure}[b]{0.19\textwidth}
         \includegraphics[width=\linewidth]{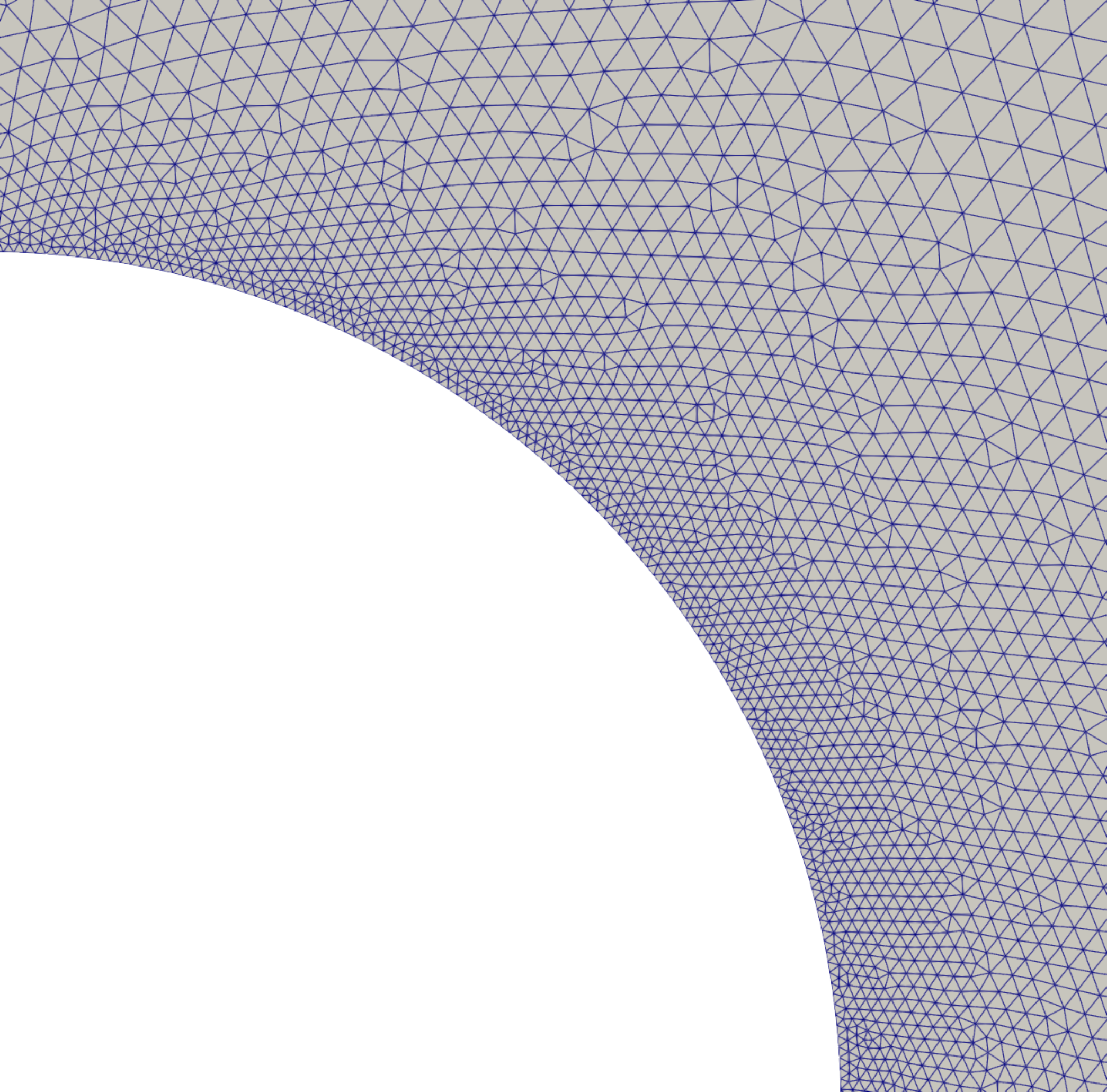}
         \caption{}
     \end{subfigure} 
    \caption{In \textbf{(a)}, a schematic of the domain of the oscillating cylinder in a free stream. In \textbf{(b)}, the computational mesh with a refined area in the wake of the cylinder. 
    The cylinder's position is based on its diameter ($D$), here placed at ($10D$, $15D$) relative to the origin. 
    We apply a constant inlet velocity of magnitude $U_\infty$ at the left, top, and bottom boundaries, and a pressure condition at the right boundary. In \textbf{(c)}, a view of the resolution at the cylinder boundary for the mesh in \textbf{(b)}.}
    \label{fig:movingcircle}
\end{figure}

\subsubsection{2D Validation – Oscillating cylinder in free stream}
For our first validation problem, we considered an oscillating cylinder in a free stream, and compared it to the sophisticated spectral element solution by Blackburn and Henderson.~\cite{blackburn1999study}
The problem consisted of a cylinder with diameter $D = 10$ cm, oscillating in a fluid flow subject to a free-stream velocity $U_\infty = 1$ m/s, as shown in Figure \ref{fig:movingcircle}\textbf{a}. 
The kinematic viscosity is set to $\nu = 2\cdot 10^{-4}$ m$^2$/s, leading to a Reynolds number of $Re = 500$, well above the critical value for vortex shedding.
Following Blackburn and Henderson, the simulation setup relied on the frequency ratio $F$ and the fixed-cylinder vortex shedding frequency $f_v$, defined as:
\begin{align}
F = \frac{f}{f_v}, \qquad f_v = \frac{St U_\infty}{D}, \label{eq:freq}
\end{align}
where $f$ is the frequency of the cylinder motion, and $St$ is the Strouhal number.
The cylinder position followed a sinusoidal profile in the transverse direction to the flow:
\begin{align}
    \b x(\b \chi, t) = (0, \hat{y}(t)) = (0, A \sin (2 \pi f t )), \label{eq:y_hat}
\end{align}
where $A = A_{ratio}D$ is the amplitude, and $A_{ratio} = 0.25$ is the amplitude ratio.
For our simulations we considered three cases where $F=0.875$, $F=0.975$ and $F=1.0$, for $St=0.228$.
At the left, top, and bottom boundaries the constant velocity $U_\infty$ was prescribed, while the right wall was considered open with $p_0 = 0$.
Following the spectral element analysis by Blackburn and Henderson, we performed a mesh convergence study based on time-dependent forces, studied the influence of the frequency ratio $F$ on the vorticity fields, and investigated cycle-convergence by comparing the cross-flow motion $\hat{y}(t)$ against the lift force \cite{blackburn1999study}.
The simulations were performed on a total of six increasingly refined computational meshes generated in Gmsh~\cite{geuzaine2009gmsh}. 
The mesh refinement level was defined in terms of the radial thickness $\Delta r$ of the innermost cell to the cylinder surface, which was approximately halved for each refinement. 
The resolution downstream of the cylinder was kept constant at $\Delta x = 0.5D$, as shown in Figure \ref{fig:movingcircle}\textbf{b}. and \ref{fig:movingcircle}\textbf{c}.
The computational domain was $\Omega_0 = [0, 62D]\times[0, 30D]$, and was refined around and in the wake region of the cylinder. 
For all simulations we used $\mathbb{P}_1/\mathbb{P}_1$ elements for velocity and pressure, respectively, and used a time step of $\Delta t = 10^{-4}$ s, keeping the Courant–Friedrichs–Lewy (CFL) number less than 0.5 to maintain numerical stability for our temporal scheme. 

\begin{figure}
    \centering
    \includegraphics[width=0.55\textwidth]{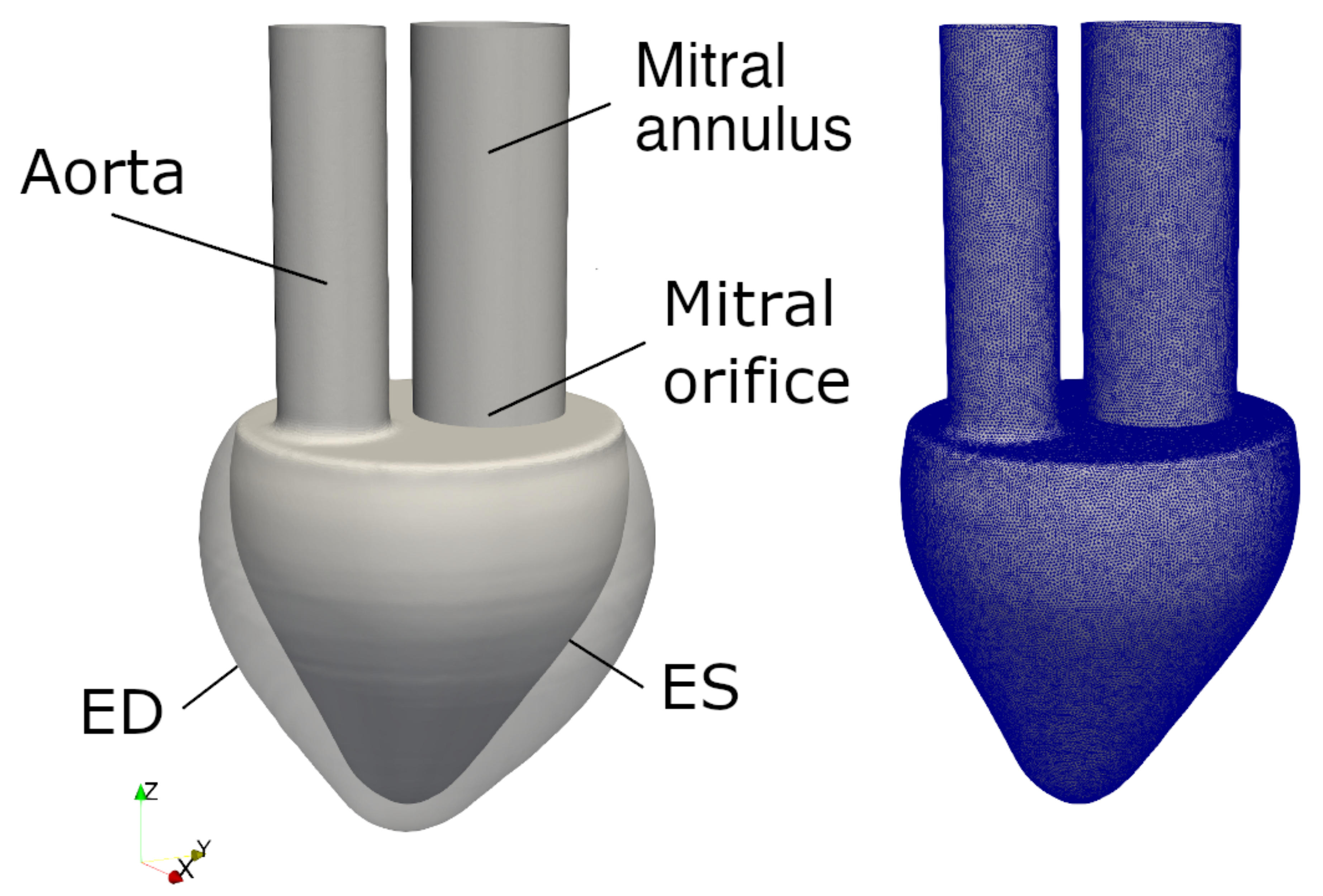} 
    \caption{The computational model employed in the current study, showing
a reconstructed left ventricle at ED and ES states on the left hand side. On the right hand side the triangulated ventricle consisting of approximately 2.5M tetrahedral cells.}
    \label{fig:ventriclemesh}
\end{figure}

\subsubsection{3D Validation – Cardiovascular flow in an idealized left ventricle}
\label{sec:3dval}
For our second validation problem, we considered three dimensional flow in a moving left ventricle (LV), and compared the results with previous numerical and experimental results by Vedula et al.~\cite{vedula2014computational}
The problem consisted of an idealized LV, generated using biplanar images from the experimental setup. 
The end-diastolic (ED) and end-systolic (ES) geometries are shown in the left panel of Figure \ref{fig:ventriclemesh}.
The computational model was scaled into non-dimensional units, and represented cardiovascular flow at a Reynolds number of $Re = 3500$ measured at the mitral orifice. 
The meshed geometry included cylindrical extensions with diameters of 0.429 and 0.304, representing the mitral and aortic orifice, respectively.
The non-dimensional viscosity was set to $\nu^* = 0.00191$, and the cardiac cycle duration was set to $T_c = 0.9$ s. 

\begin{figure}
     \centering
     \begin{subfigure}[b]{0.35\textwidth}
         \centering
         \includegraphics[width=\textwidth]{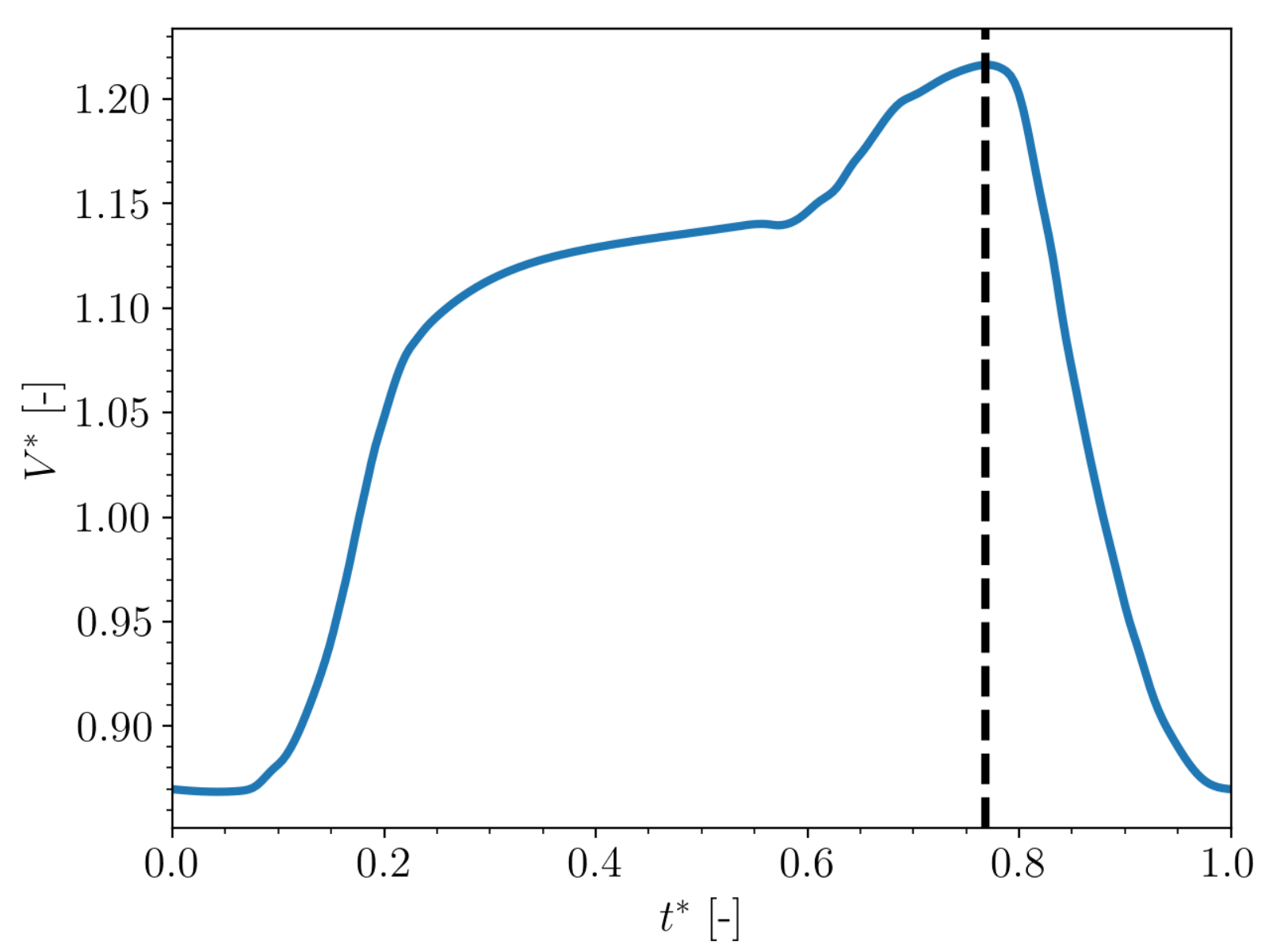}
         \caption{}
     \end{subfigure}
     \begin{subfigure}[b]{0.35\textwidth}
         \centering
         \includegraphics[width=\textwidth]{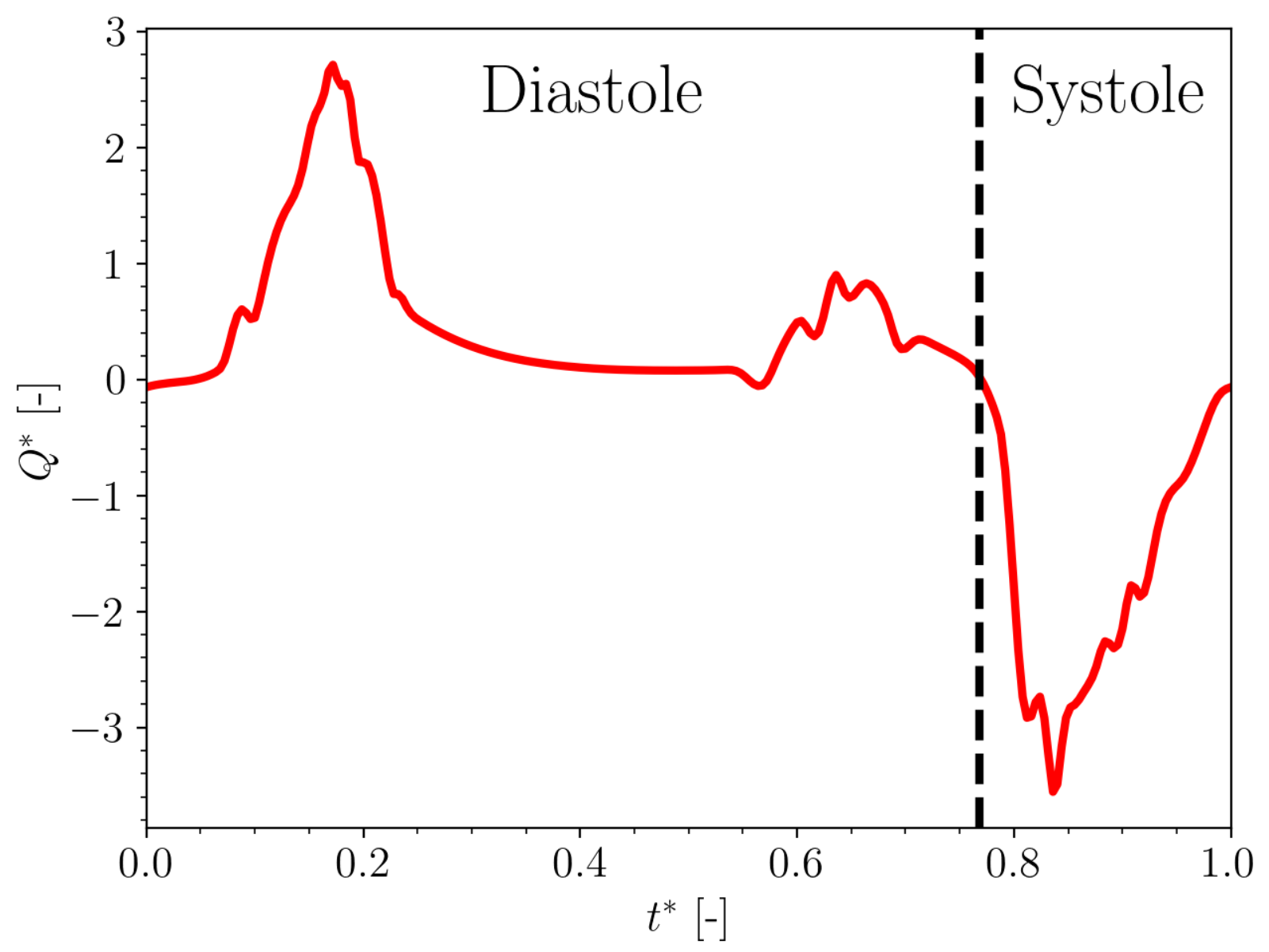}
         \caption{}
     \end{subfigure} 
  \caption{In \textbf{(a)}, the non-dimensional volume ($V$) of the LV computed directly from the displaced mesh. In \textbf{(b)}, the corresponding non-dimensional boundary flow rate ($Q$) computed from the LV volume curve, used to prescribe the inlet velocity profile at the mitral valve. The dashed line represents the transition between the ventricular diastolic and systolic phases.}
  \label{fig:vedula-vol-rate}
\end{figure}

\begin{figure}
     \centering
     \begin{subfigure}[b]{0.20\textwidth}
         \includegraphics[width=\linewidth]{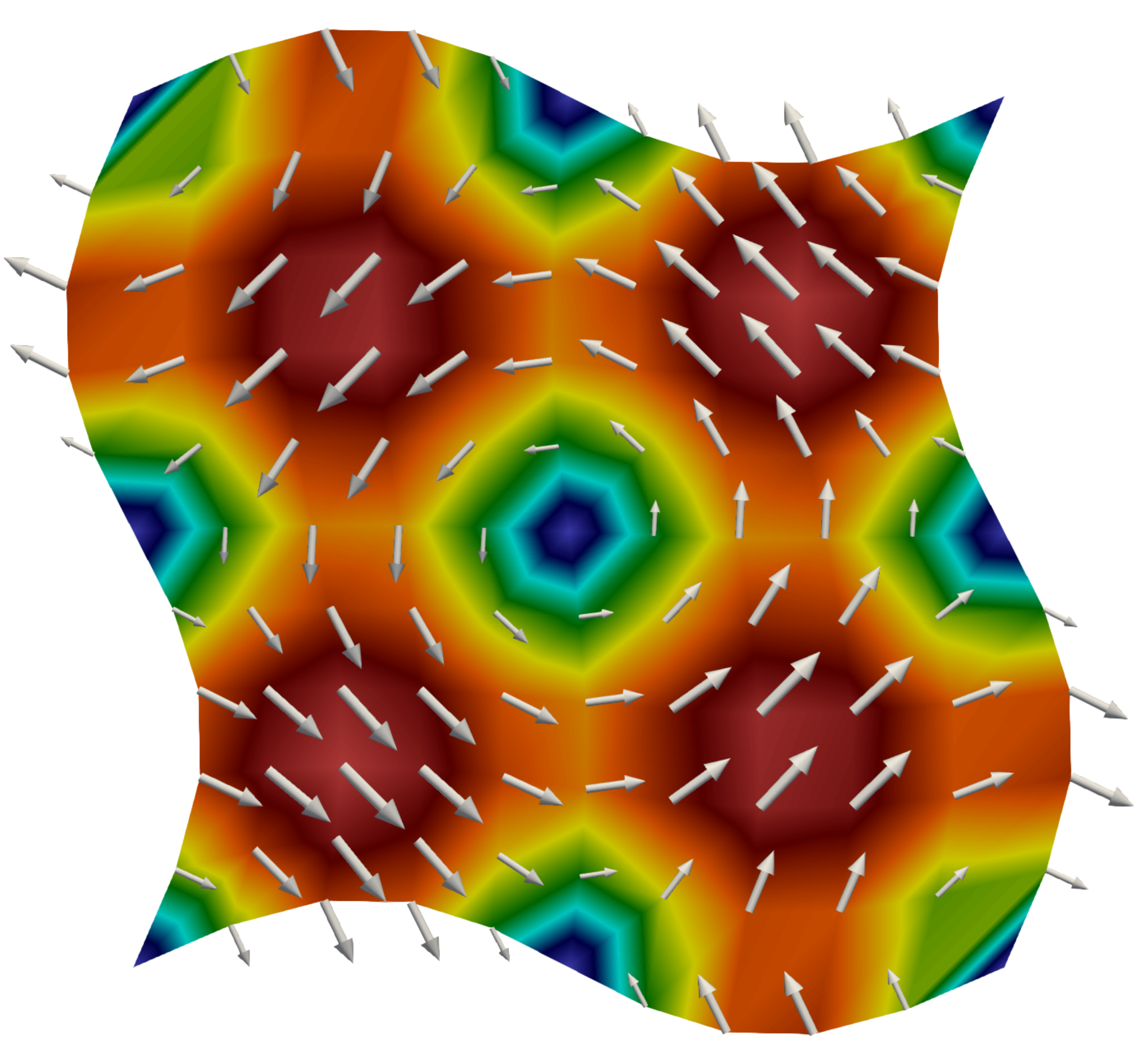}
         \caption{}
     \end{subfigure}
     \begin{subfigure}[b]{0.20\textwidth}
         \includegraphics[width=\linewidth]{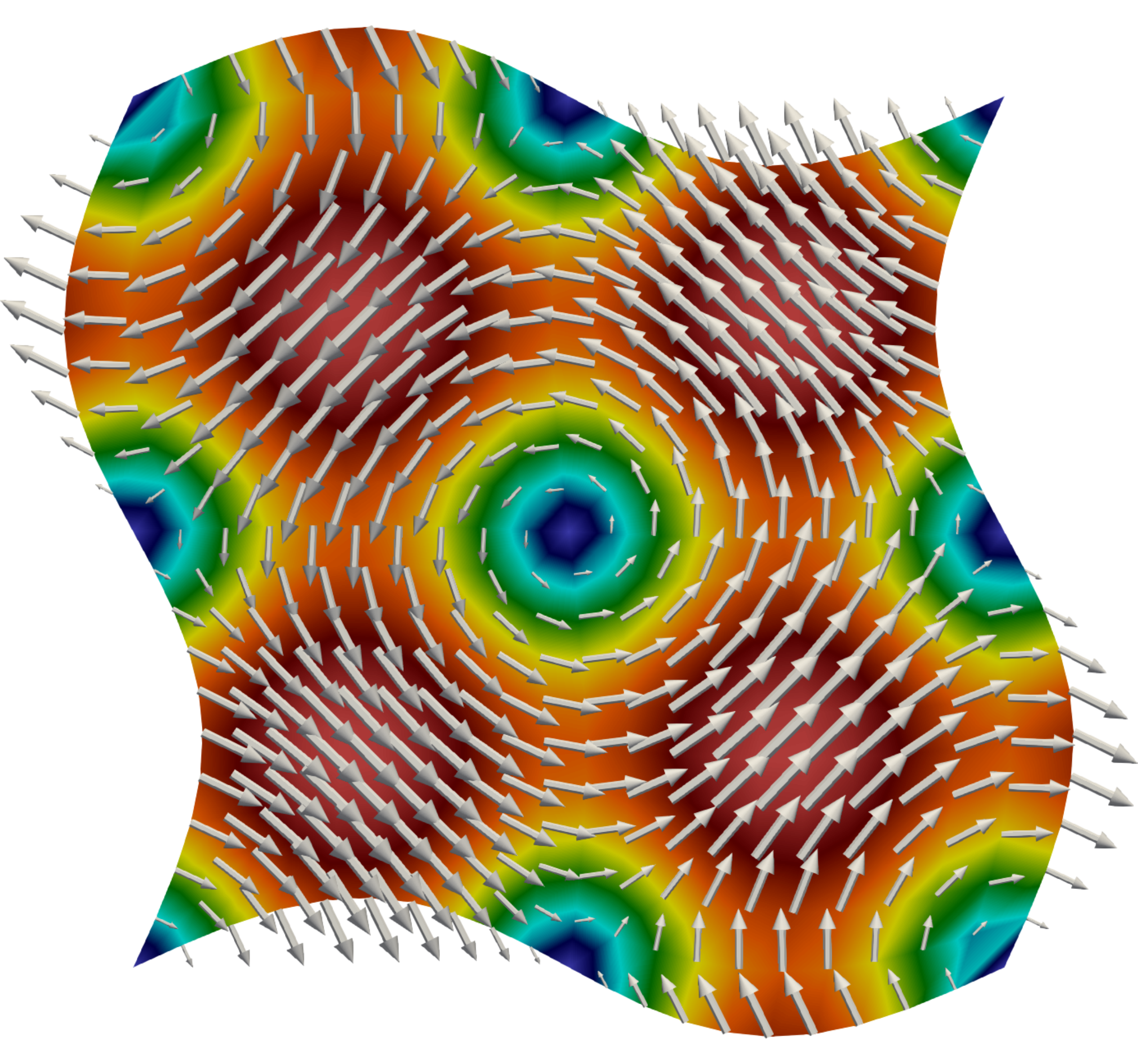}
         \caption{}
     \end{subfigure} 
     \begin{subfigure}[b]{0.20\textwidth}
         \includegraphics[width=\linewidth]{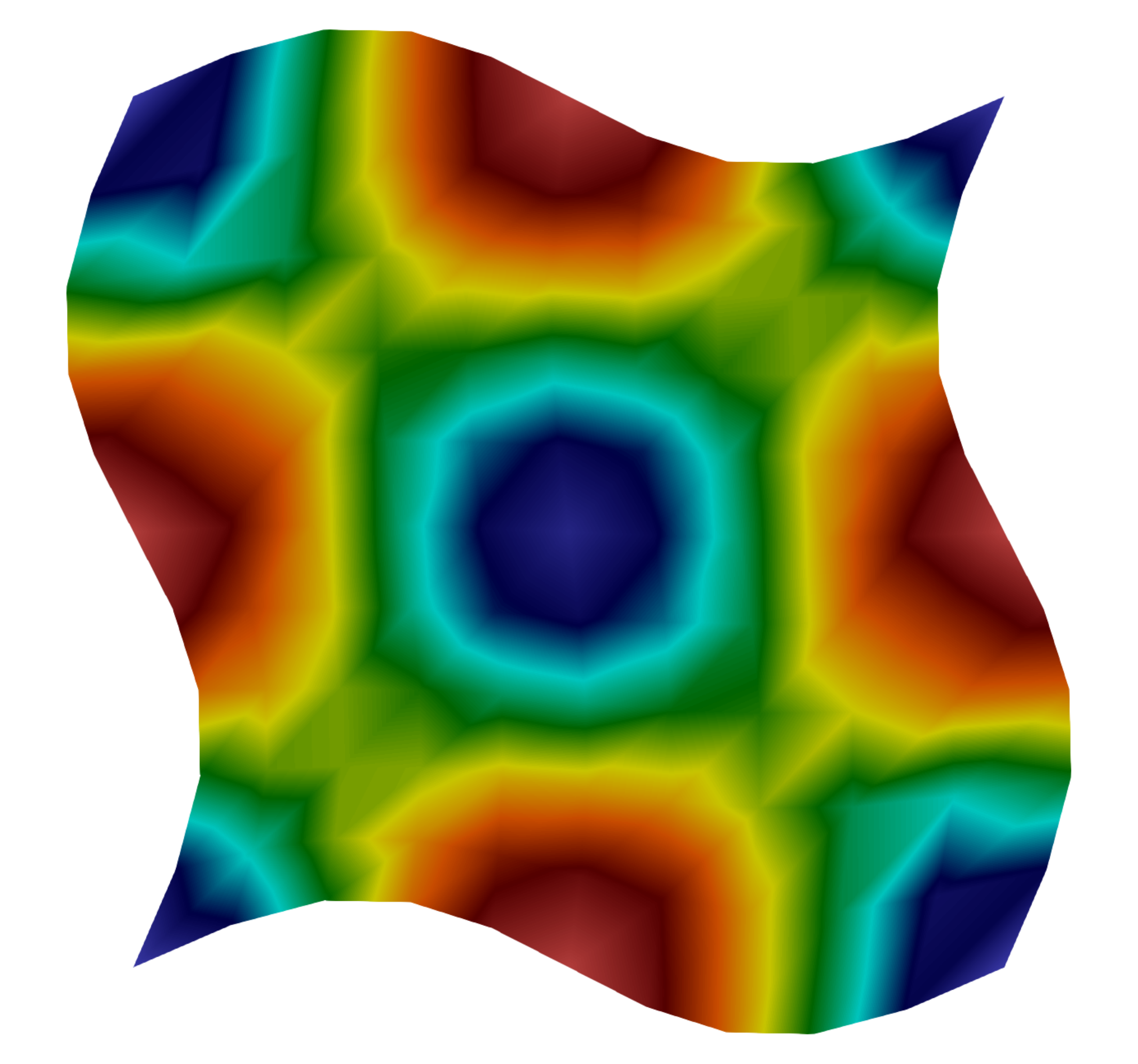}
         \caption{}
     \end{subfigure} 
     \begin{subfigure}[b]{0.20\textwidth}
         \includegraphics[width=\linewidth]{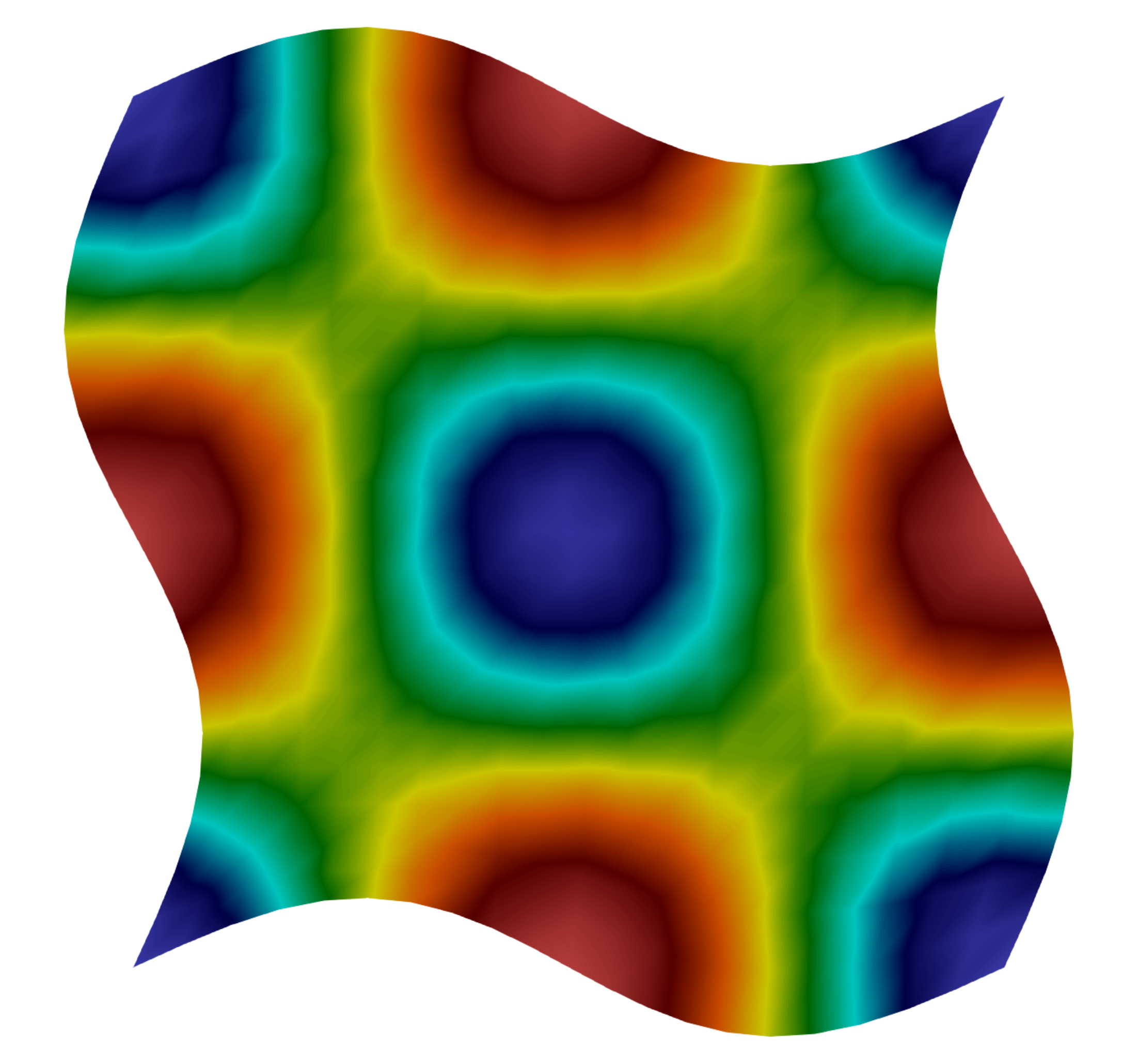}
         \caption{}
     \end{subfigure} 
    \caption{Instantaneous velocity and pressure field at $t = T = 1.0$ for the vortex problem. 
    The mesh resolution varies between 200 and 800 cells, for the results shown in \textbf{(a)} and \textbf{(c)}, and \textbf{(b)} and \textbf{(d)}, respectively. 
    Both simulation were run using $\mathbb{P}_2$/$\mathbb{P}_1$ finite elements for velocity and pressure, and with mesh deformation amplitude of $A=0.08$.} 
    \label{fig:mms-ale-viz}
\end{figure}

In contrast to the previous problems, there was no analytic expression for the LV's boundary motion.  
Consequently, a set of $N=126$ surface models were generated from the idealized silicone model, reflecting one cardiac cycle.\footnote{The 126 surface models were obtained through personal communication with the authors of the paper by Vedula et al.\cite{vedula2014computational}}
Starting with one template surface model, a volumetric mesh consisting of approximately 7M tetrahedral cells was generated using the \textit{Vascular Modeling Toolkit} (VMTK) software ~\cite{antiga2008image}.
The remaining $N - 1$ surfaces models were then projected onto the mesh to calculate the respective displacement fields, $\b x(\b \chi,t)$.  
However, since the LV motion was needed at several time steps throughout the cardiac cycle, and not only at times $t_0,t_1,\cdots,t_{N-1}$, we created a continuous mapping of the deformation and mesh velocity between each time step. 
This was achieved by performing B-spline approximation using the \texttt{splrep} method in Scipy~\cite{virtanen2020scipy} to compute the mesh displacement at each node.
The first order derivative of the resulting spline was then computed to obtain an expression for the mesh velocity $\b w(\b \chi, t)$ on the boundary, which was applied as the ventricle wall boundary condition.
For this problem, we prescribed a velocity profile at the mitral valve (MV), and
a pressure condition at the aortic valve (AV), working as the inlet and outlet,
respectively. The prescribed velocity profile ensured that the flow entered
through the MV and exited through the AV, and to avoid a downwind scheme effect,
which is known to be unconditionally unstable~\cite{sewell2005numerical}. 
The velocity profile to be prescribed at the MV was derived by considering conservation of mass in the domain:
\begin{align}
    Q_{\text{MV}} + Q_{\text{AV}} + \frac{\d V(t)}{\d t} = 0, \label{eq:qmv}
\end{align}
where $Q_{MV}$ and $Q_{AV}$ are the respective volumetric flow rates, and $V$ denotes the LV volume.
Because the deformation was known, we could compute the volume change over one cardiac cycle, shown in Figure \ref{fig:vedula-vol-rate}\textbf{a}.
From the volume curve, we obtained the corresponding "boundary" flow rate from its first order derivative, shown in Figure \ref{fig:vedula-vol-rate}\textbf{b}.
Assuming that the AV is closed during ventricular diastole, i.e., $Q_{\text{AV}} = 0$, we could deduce the average velocity through the MV from \eqref{eq:qmv}:
\begin{align}
      Q_{\text{MV}} = \bar{\b u}_{\text{MV}} A_{\text{MV}}  = - \frac{\d V(t)}{\d t} \quad \Longleftrightarrow \quad \bar{\b u}_{\text{MV}}(t)  = -\frac{1}{A_{\text{MV}}}\frac{\d V(t)}{\d t},
\end{align}
where $A_{\text{MV}}$ is the MV's cross-sectional area.
Based on this analysis, a plug velocity profile was prescribed at the MV during diastole, with average velocity $\bar{\b u}_{\text{MV}}(t)$. 
The AV was considered open with $p_0 = 0$, but the aortic flow was effectively zero since the prescribed flow rate through the MV was equal to the total volume change of the LV. 
Similarly, during systole, the MV flow rate was set to zero while the AV boundary condition remained the same, ensuring that the flow exited through the AV. 

To ensure adequate spatial accuracy, we performed a mesh refinement study using four computational meshes of varying resolution: coarse (600K cells), medium (2.5M cells), fine (7M cells), and very fine (20M cells). 
Similarly, to ensure that non-physiologic effects from initialization did not affect the solution, we performed a cycle convergence test, where we ran one simulation over 10 cardiac cycles.
For both the mesh and cycle convergence tests, the time step was kept constant to minimize the influence of temporal discretization errors.
A time step of $\Delta t = 2\cdot 10^{-4} T_c$ and $\Delta t = 4\cdot 10^{-4}T_c$ was used for the mesh and cycle convergence, respectively, assuring a mean CFL number less than 0.2. 

\begin{figure}[b]
\centering
    \begin{subfigure}{.45\textwidth}
        \centering
        \includegraphics[width=0.9\textwidth]{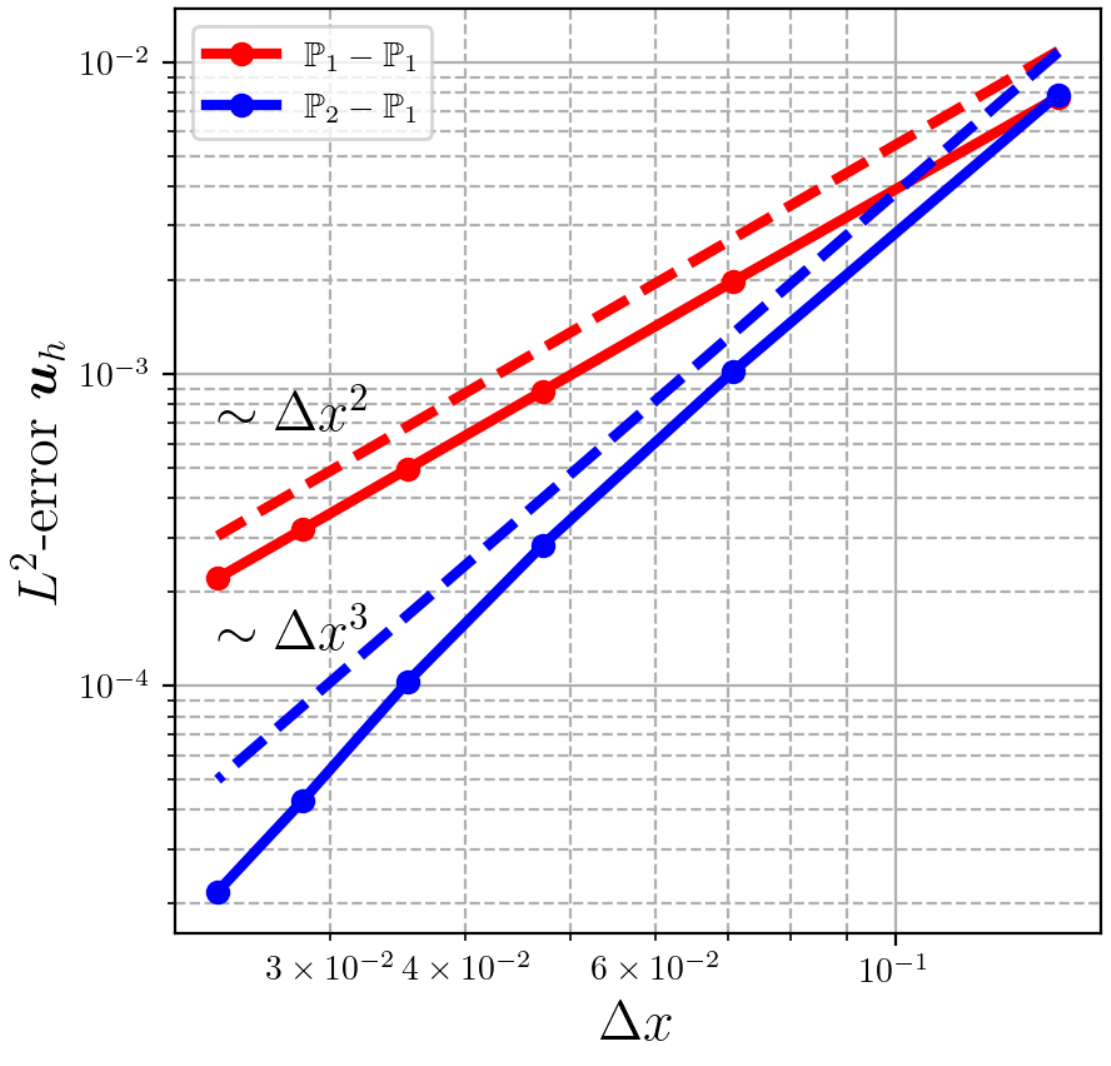}
        \caption{}
    \end{subfigure}%
    \begin{subfigure}{.45\textwidth}
        \centering
        \includegraphics[width=0.9\textwidth]{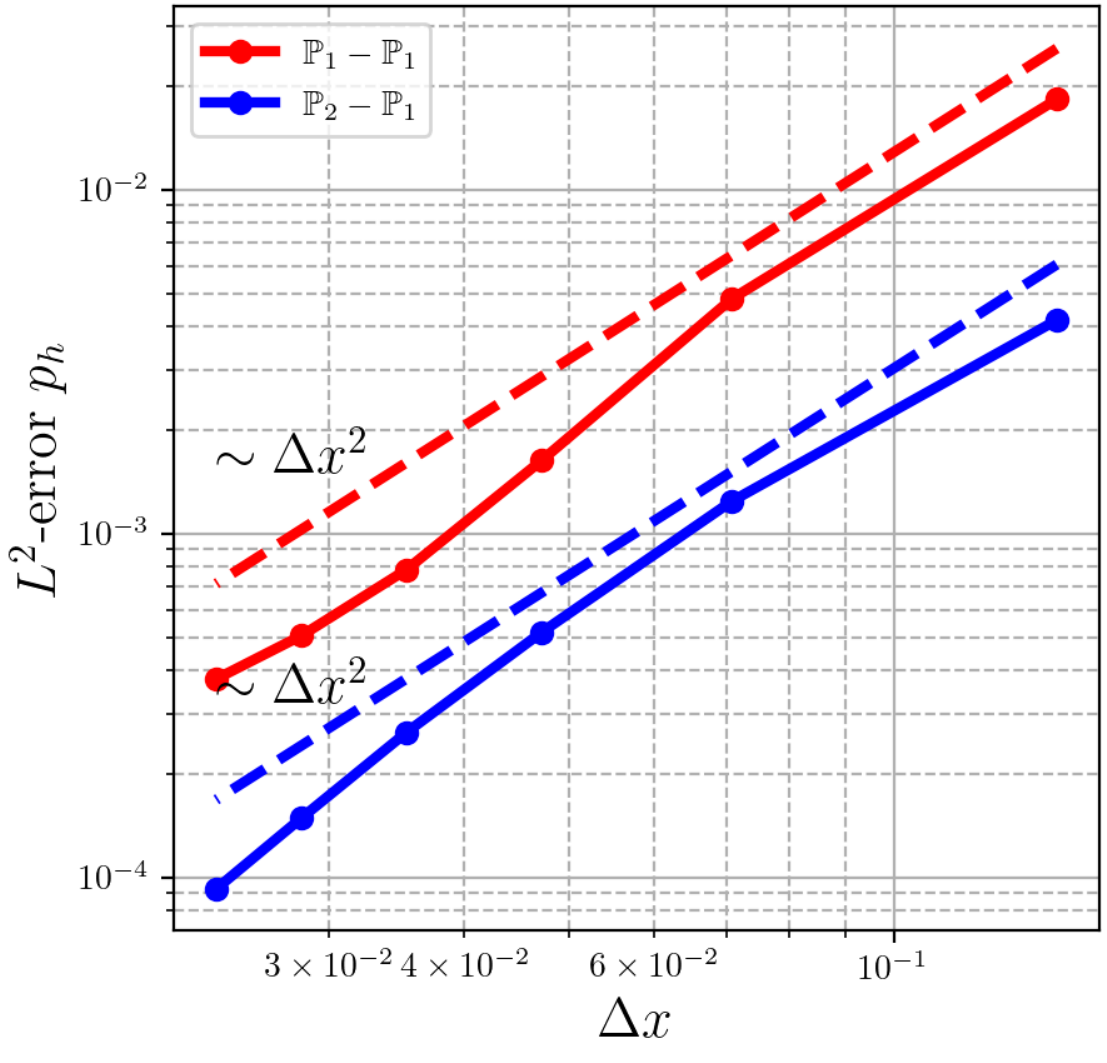}
        \caption{}
    \end{subfigure}
  \caption{Spatial convergence study performed by varying the characteristic edge length $\Delta x$.
  In \textbf{(a)}, the $L^2$ error for the velocity, and in \textbf{(b)} the $L^2$ error for the pressure. 
  The solid lines represent the simulation results, and the dashed lines show the theoretical convergence rates. 
  }
   \label{fig:mms-ale-dx}
\end{figure}

\begin{figure}[b]
\centering
    \begin{subfigure}{.45\textwidth}
        \centering
        \includegraphics[width=0.9\textwidth]{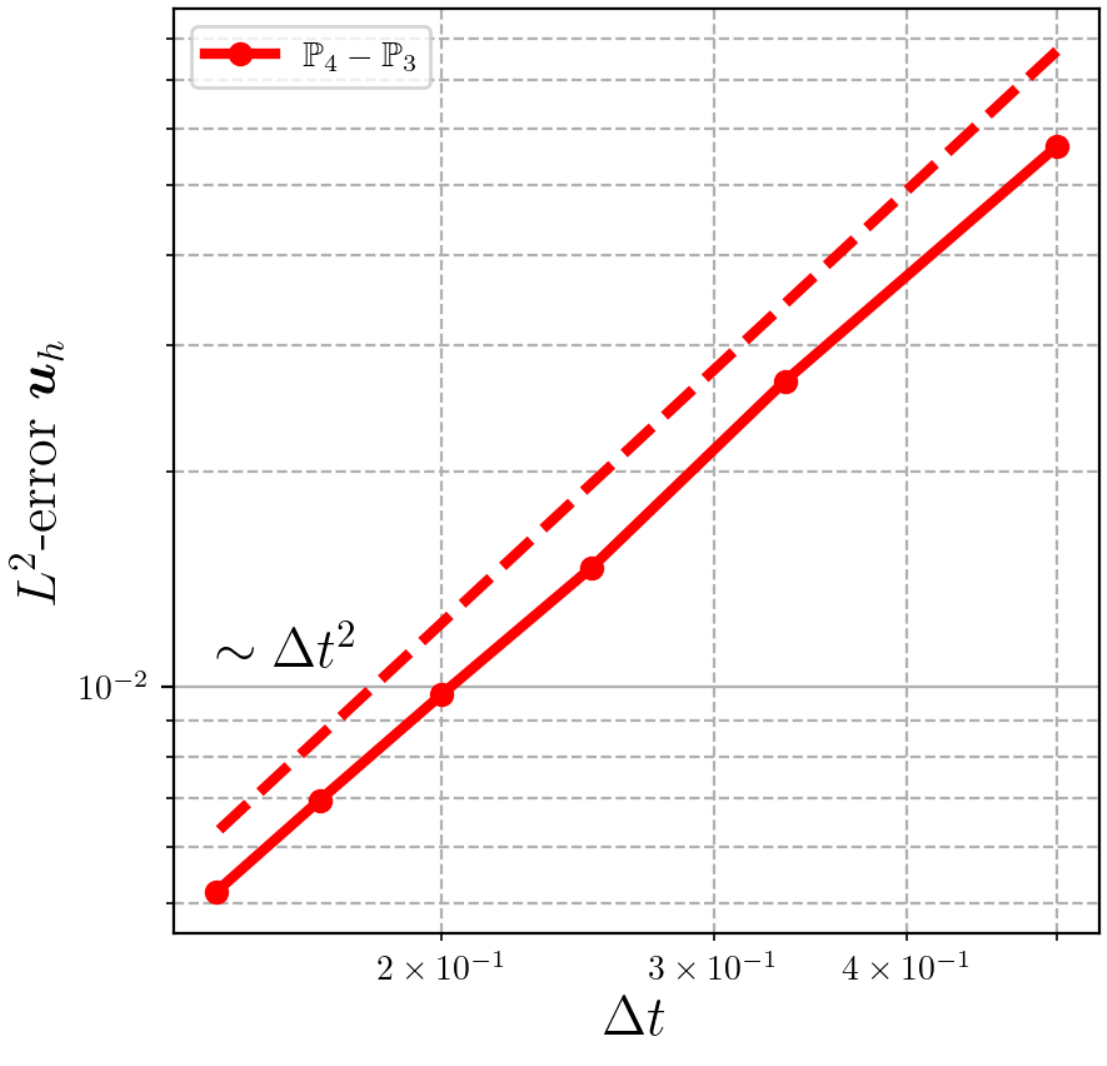}
        \caption{}
    \end{subfigure}%
    \begin{subfigure}{.45\textwidth}
        \centering
        \includegraphics[width=0.9\textwidth]{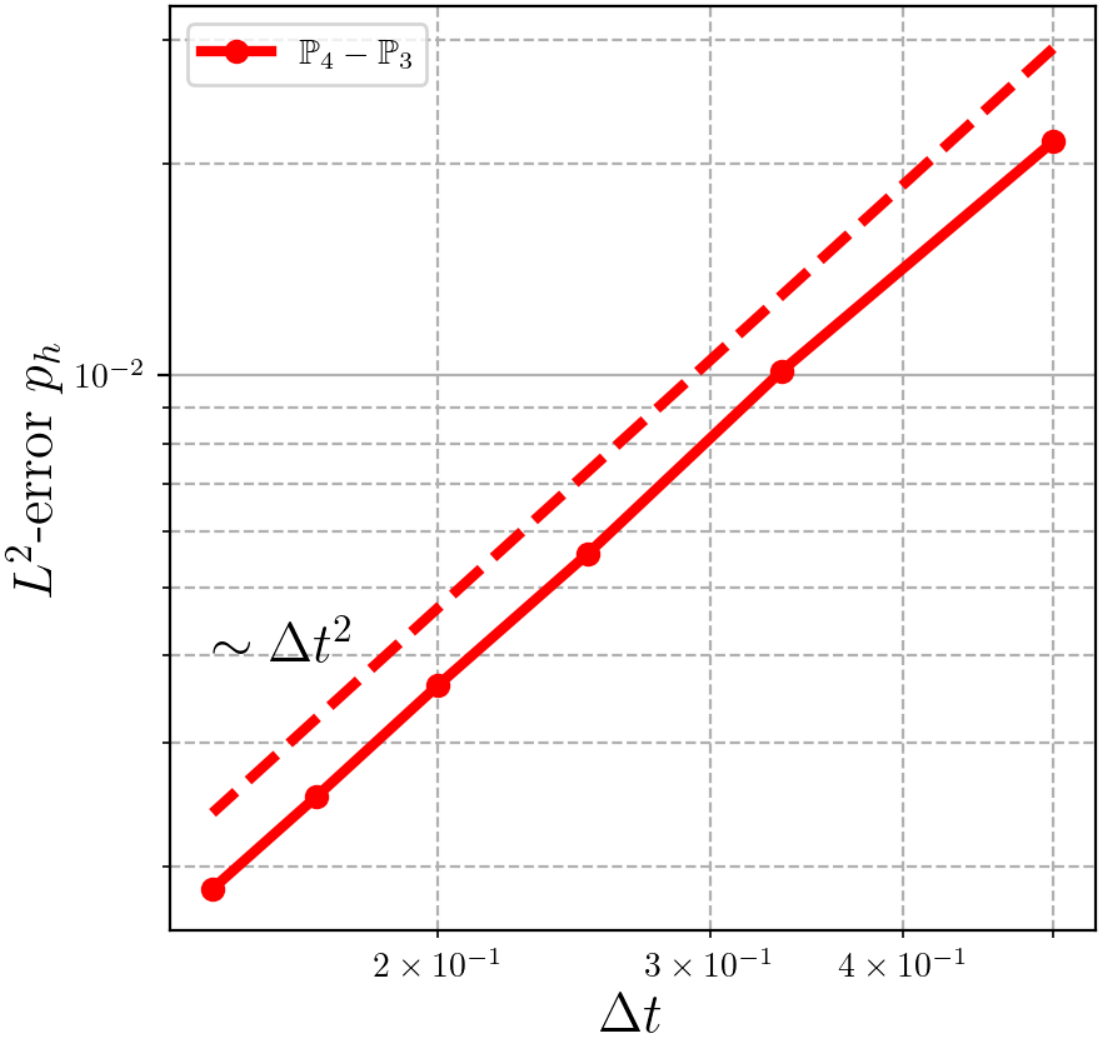}
        \caption{}
    \end{subfigure}
  \caption{Temporal convergence study performed by varying the time step $\Delta t$.
   In \textbf{(a)}, the $L^2$ error for the velocity, and in \textbf{(b)} the $L^2$ error for the pressure. 
  The solid lines represent the simulation results, and the dashed lines show the theoretical convergence rates.
  }
  \label{fig:mms-ale-dt}
\end{figure}

\section{Results}
\label{results}
\subsection{2D Verification – Vortex problem with oscillating boundary}
Figure \ref{fig:mms-ale-viz} shows the solution, in terms of velocity magnitude and pressure, at the time of maximal mesh deformation, for two different mesh resolutions. 
We also present a vector representation of the velocity solution, scaled by the magnitude.  
Velocity for the coarse mesh is shown in Figure \ref{fig:mms-ale-viz}\textbf{a}, and the difference is qualitatively negligible compared to the velocity field for the finer mesh in Figure \ref{fig:mms-ale-viz}\textbf{b}. 
The pressure field for the coarse mesh in Figure \ref{fig:mms-ale-viz}\textbf{c} displays non-smoothness in the transitional regions between the vortices, whereas the finer mesh in Figure \ref{fig:mms-ale-viz}\textbf{d} gives a visibly smoother pressure solution. 

\begin{table}
    \centering
    \begin{tabular}{rrrrrr}
    \hline
    \hline
   $\Delta r$ &   $N_{cells}$ &   $\#dofs$ &   $\hat C_L$ &   $\bar C_D $ &   $-\bar C_{pb}$ \\
\hline
        $4.5 \cdot 10^{-3}$ &         13893 &       7084 &        1.273 &         1.547 &            1.053  \\
        $2.2 \cdot 10^{-3}$ &         15196 &       7767 &        1.787 &         1.387 &            1.360  \\
        $1.0 \cdot 10^{-3}$  &         20218 &      10341 &        1.752 &         1.400 &            1.309  \\
        $5.0 \cdot 10^{-4}$  &         34270 &      17493 &        1.766 &         1.411 &            1.312  \\
        $2.5 \cdot 10^{-4}$  &         66620 &      33920 &        1.768 &         1.412 &            1.304  \\
        $1.3 \cdot 10^{-4}$  &        162871 &      82548 &        1.772 &         1.414 &            1.318  \\

    \hline
            -   & 422   & 23100   &  \textbf{1.776}   & \textbf{1.414}   & \textbf{1.377}      \\
    \hline
    \end{tabular}
     \caption{Mesh convergence results for the oscillating cylinder, where $A = 0.25$, and $F = 1.0$. 
     The table displays the radial edge length ($\Delta r$), number of cells in the mesh ($N_{cells})$, number of degrees of freedom ($\# dofs$), peak lift coefficient ($\hat{C_L}$), mean drag coefficient ($\bar{C_D}$), and the mean pressure coefficient ($\bar{C_{pb}}$). 
     Reference data by Blackburn and Henderson has been emphasized in bold font.}
     \label{tab:cylinder-moving}
\end{table}

\begin{figure}
    \centering
     \includegraphics[width=0.35\textwidth, angle=-90]{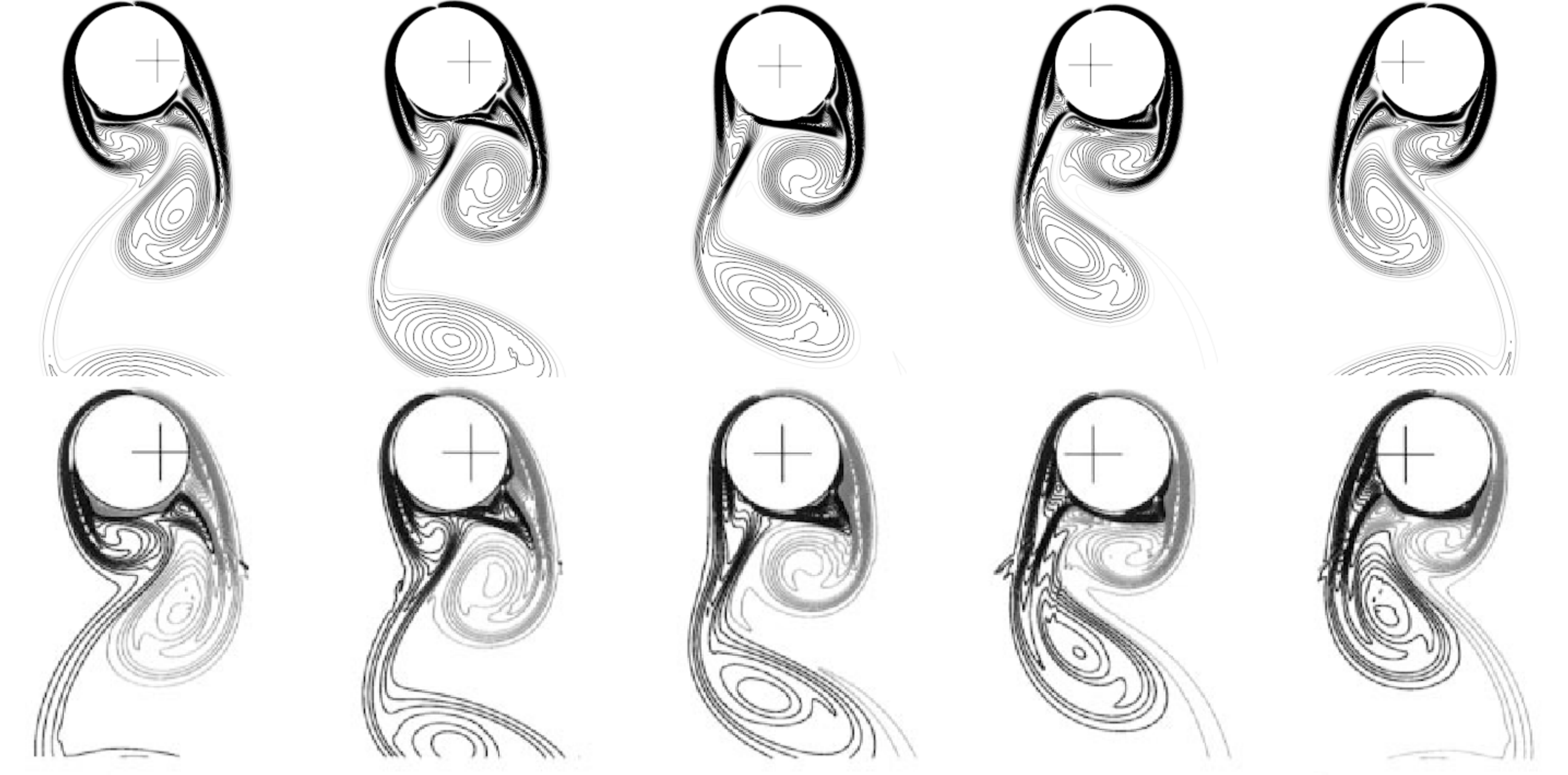}
    \caption{Instantaneous vorticity contours ranging over half a period of oscillation, for $F=0.0975$ and $A_{ratio}=0.25$.
    Present simulation results (top row) are compared with reference data by Blackburn and Henderson~\cite{blackburn1999study} (bottom row) at five snapshots throughout the half period. Note that the figure has been rotated 90 degrees clockwise for visualization purposes.}
    \label{fig:cylinder-vortex}
\end{figure}

Results from the spatial and temporal convergence tests are presented in Figures \ref{fig:mms-ale-dx} and \ref{fig:mms-ale-dt}, respectively, where the dashed lines indicate the theoretical order of convergence in the $L^2$ norm.
The theoretical spatial and temporal convergence rates, $p$ and $q$, can be derived from the $L^2$ error of the solution: 
\begin{align}
    E = \alpha \Delta x^p + \beta \Delta t^q, \label{eq:error}
\end{align}
where $\alpha$ and $\beta$ are independent constants, $\Delta x$ is the characteristic edge length, and $\Delta t$ is the time step.
To determine $p$, the temporal discretization error must be negligible compared to the spatial error, $ \alpha \Delta x^q  \gg \beta \Delta t^q $. 
Thus, as $\Delta t \rightarrow 0$, the error becomes mainly dependent on the mesh resolution, $E \approx \alpha \Delta x^p$, and the convergence rate can be computed as follows:
\begin{align}
    p = \frac{\log\left(\frac{E_{k+1}}{E_k}\right)}{\log\left(\frac{\Delta x_{k+1}}{\Delta x_{k}}\right)}, \label{eq:convergence}
\end{align}
where $E_{k+1}$ is computed on a more refined mesh compared to $E_k$.
Temporal convergence rates were computed in the same manner, by letting $\alpha \Delta x^p \ll \beta \Delta t^q$, and using higher order finite elements.
Focusing on spatial convergence for the velocity in Figure \ref{fig:mms-ale-dx}\textbf{a}, the results indicate that using $\mathbb{P}_1$ and $\mathbb{P}_2$ elements result in a second and third order accuracy, respectively. 
Figure \ref{fig:mms-ale-dx}\textbf{b} presents spatial convergence results for pressure, and shows that the solution is second order accurate for both element combinations. 
In Figures \ref{fig:mms-ale-dt}\textbf{a} and \textbf{b}, we observe that the temporal convergence is second order for both velocity and pressure. 

\subsection{2D Validation – Oscillating cylinder in free stream}
Mesh convergence results for frequency ratio of $F=1.0$ from \eqref{eq:freq} are presented in Table \ref{tab:cylinder-moving}, which shows global flow coefficients computed with respect to the position of the cylinder. 
Specifically, we present the mean drag coefficient $\bar{C_D}$, the peak lift coefficient $\hat{C_L}$, and the mean pressure coefficient $-\bar{C_{pb}}$, defined as:
\begin{equation}
    C_D = \frac{2 F_D}{\rho U_\infty^2 D}, \qquad C_L = \frac{2F_L}{\rho U_\infty^2 D}, \qquad C_{pb} = 1 + \frac{2(p_{180} - p_0)}{\rho U_\infty^2}.
\end{equation}
Here, $F_D$ and $F_L$ are the drag and lift force, respectively, and $p_0$ and $p_{180}$ are the pressures at the furthest upstream and downstream points on the cylinder surface.
As the mesh is refined, all three coefficients are shown to converge towards the reference values, which have been highlighted in bold. 

Second, a visualization of vorticity contours is presented in Figure \ref{fig:cylinder-vortex}, for $F=0.975$.
The present simulation results (top row) are compared to the vorticity contours of Blackburn and Henderson~\cite{blackburn1999study}. 
A qualitative comparison between the results strongly indicate that our solver is able to accurately predict and capture both the spatial dynamics of the vortex shedding and the temporal evolution of the reattachment points.

\begin{figure}
    \centering
    \includegraphics[width=0.55\textwidth]{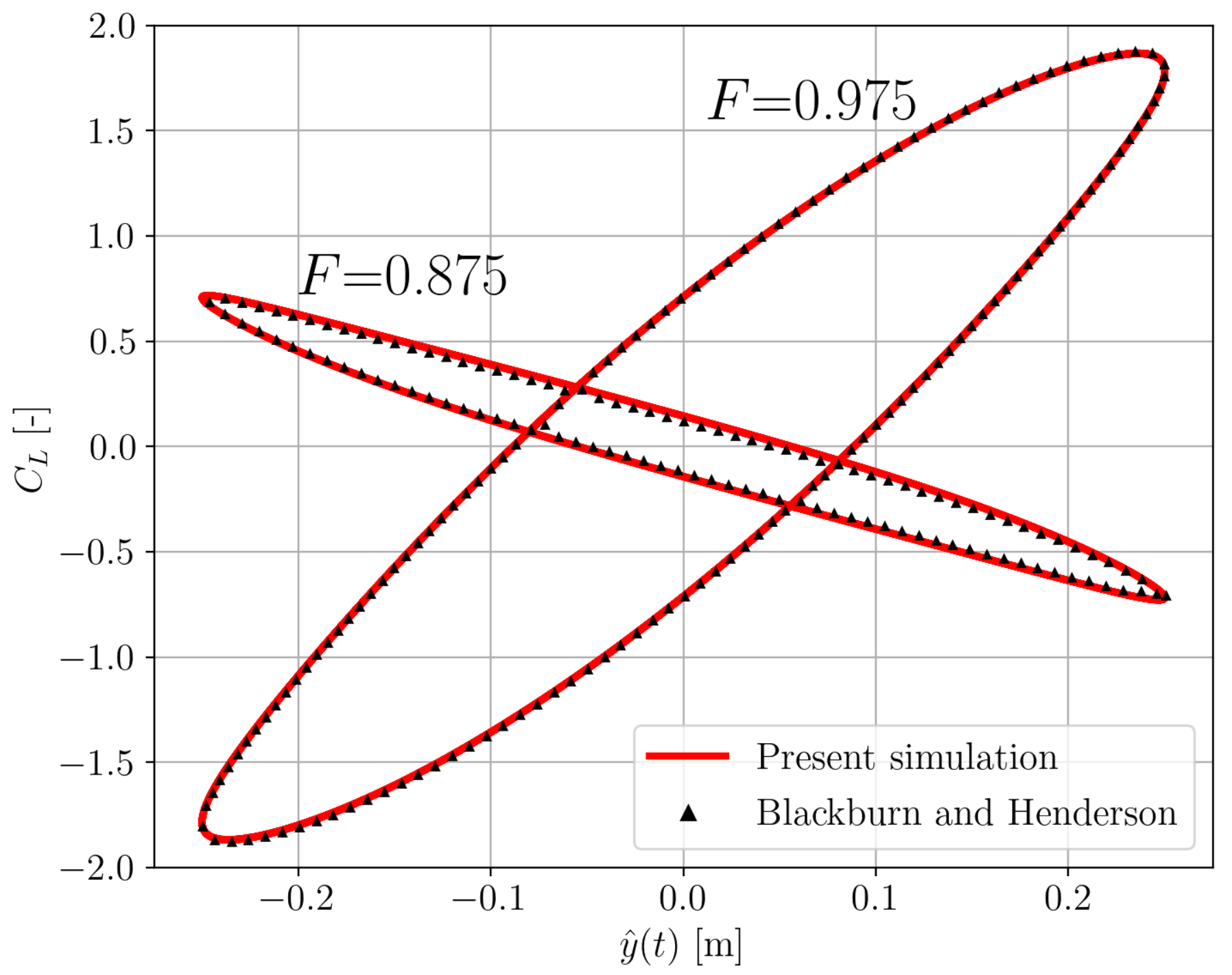} 
    \caption{
    Lift coefficient $C_L$ vs. cross-flow displacement of the cylinder $\hat{y}(t)$ for the flow past an oscillating cylinder, for frequency ratios $F=0.975$, and $F=0.0875$, over the 10 last periods of oscillations. The present simulation results (red curves) have been compared with reference data by Blackburn and Henderson (black markers).}
    \label{fig:cylinder-phase}
\end{figure}

Finally, we present the evolution of the lift coefficient $C_L$, as a function of the cylinder displacement $\hat{y}(t)$, in Figure \ref{fig:cylinder-phase}, over the 10 last periods of oscillations for frequency ratios of $F=0.875$ and $F=0.975$. 
The results are observed to match well with the reference data of Blackburn and Henderson, shown as
black markers in the figure. 

\subsection{3D Validation – Cardiovascular flow in an idealized left ventricle}
\begin{figure}
\centering
\begin{subfigure}[b]{.45\linewidth}
\includegraphics[width=\linewidth]{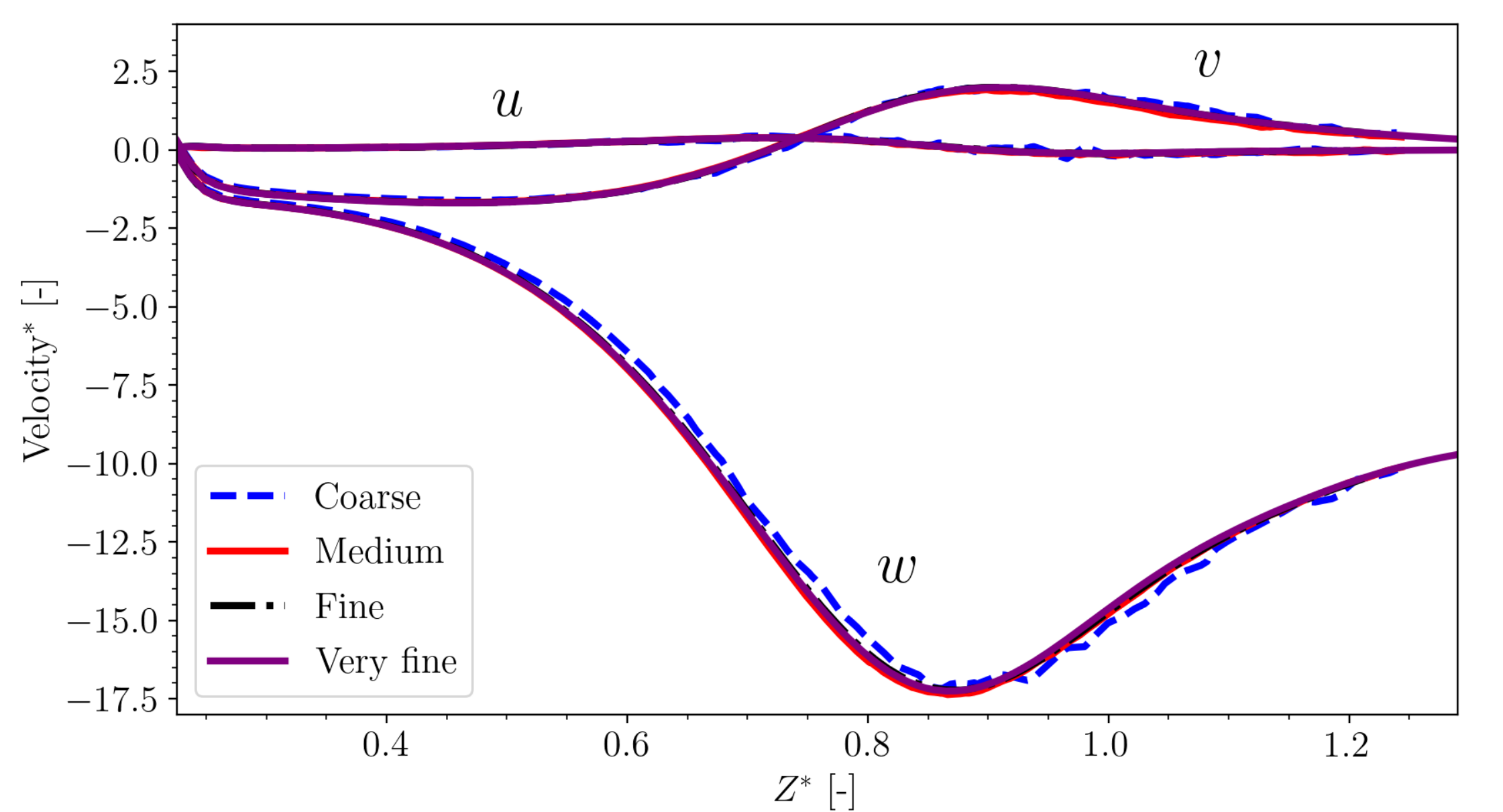}
\caption{}\label{fig:vedula-conv-u}
\end{subfigure}
\begin{subfigure}[b]{.45\linewidth}
\includegraphics[width=\linewidth]{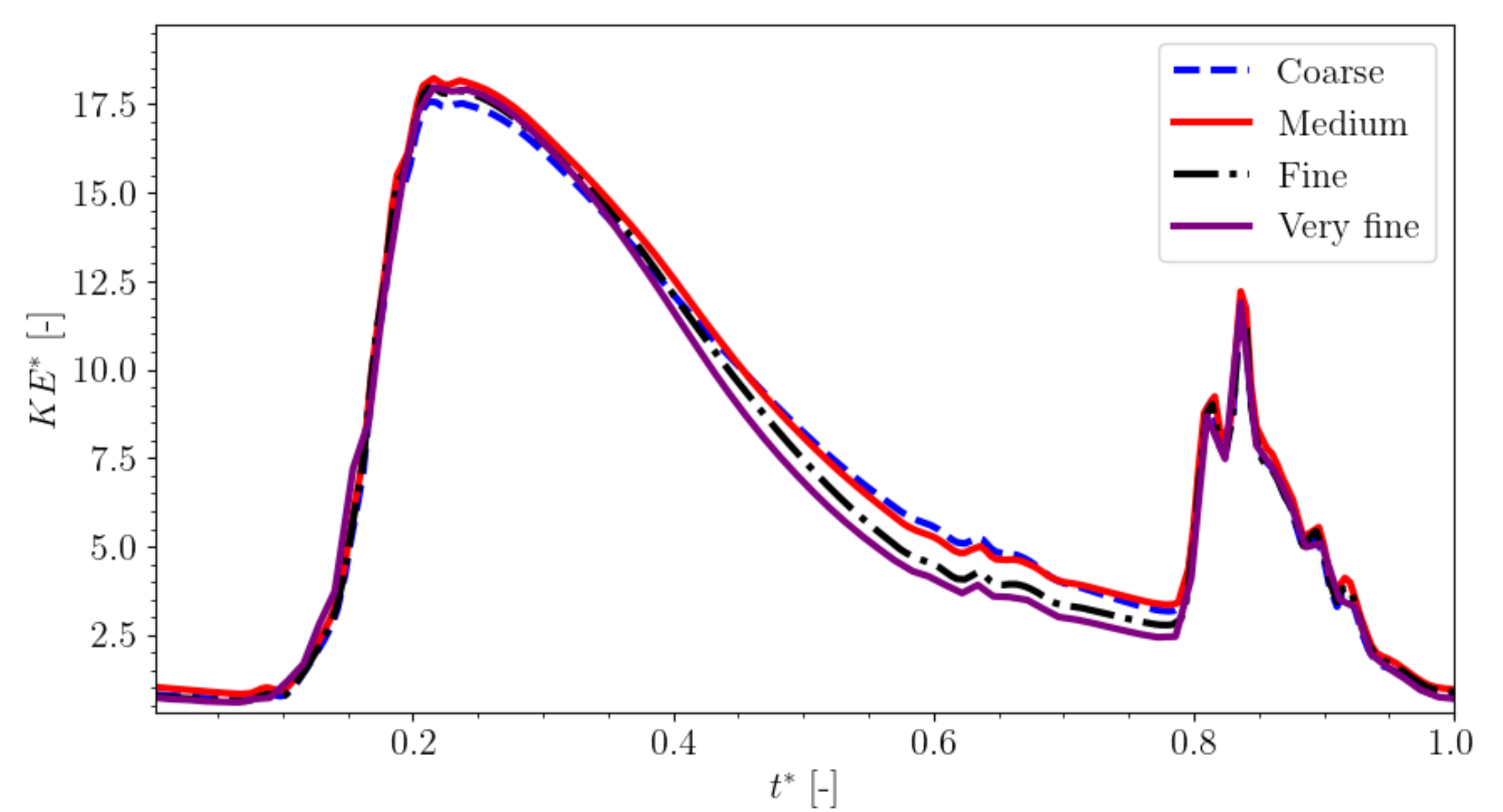}
\caption{}\label{fig:vedula-conv-ke}
\end{subfigure}

\begin{subfigure}[b]{.95\linewidth}
\includegraphics[width=\linewidth]{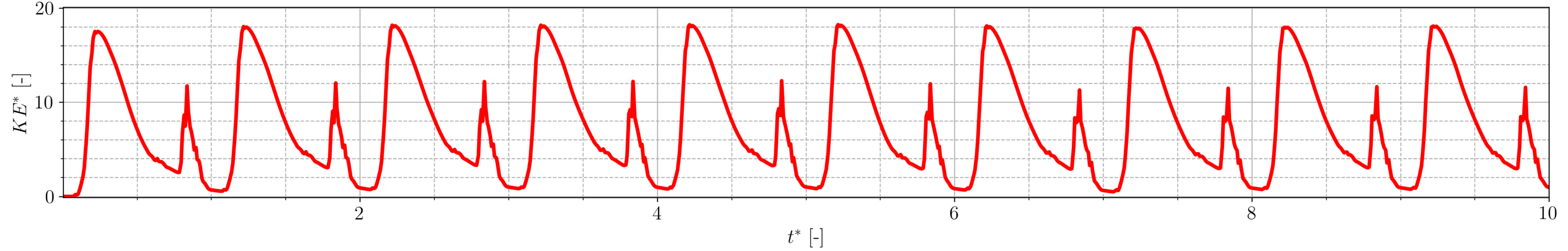}
\caption{ }\label{fig:vedula-conv-cycle}
\end{subfigure}

\caption{In \textbf{(a)} and \textbf{(b)}, results from the mesh refinement study for four resolutions: coarse (800k cells), medium (2.5M cells), fine (7M cells), and very fine (20M cells). In \textbf{(a)}, a comparison of velocity components along the mitral valve centerline at $t^* = 0.22$. In \textbf{(b)}, temporal evolution of the total kinetic energy inside the LV cavity. In \textbf{(c)}, the cycle to cycle variation in the total kinetic energy over 10 cardiac cycles.}
\label{fig:vedula-convergence}
\end{figure}

For the following results we use the non-dimensional time, defined as $t^* = t / T_c$. 
Results from the mesh refinement study on the LV geometry are shown in Figure \ref{fig:vedula-convergence}.  
Panel \textbf{a} displays a comparison of the three velocity components along the vertical line passing through the middle of the mitral valve, during the first deceleration phase at $t^* = 0.22$, while panel \textbf{b} shows the total non-dimensional kinetic energy ($KE^*$) of the fluid in the LV cavity, defined as:
\begin{equation}
    KE^* = \frac{1}{2}(u^2 + v^2 + w^2).
\end{equation}
A reasonable convergence is achieved as the mesh is refined, with a noticeably smoother velocity profile, particularly the $w$ component, and a gradual decrease in kinetic energy inside the cavity.
There are negligible differences in the velocity profiles between the 2.5M, 7M, and 20M cell solutions. 
The results from the cycle convergence study are shown in Figure \ref{fig:vedula-convergence}\textbf{c}, displaying the $KE^*$ over ten cardiac cycles.
The results show no noticeable difference in neither $KE^*$ shape nor magnitude, with exception of the first cycle.

\begin{figure}
    \centering
    \includegraphics[width=0.75\textwidth]{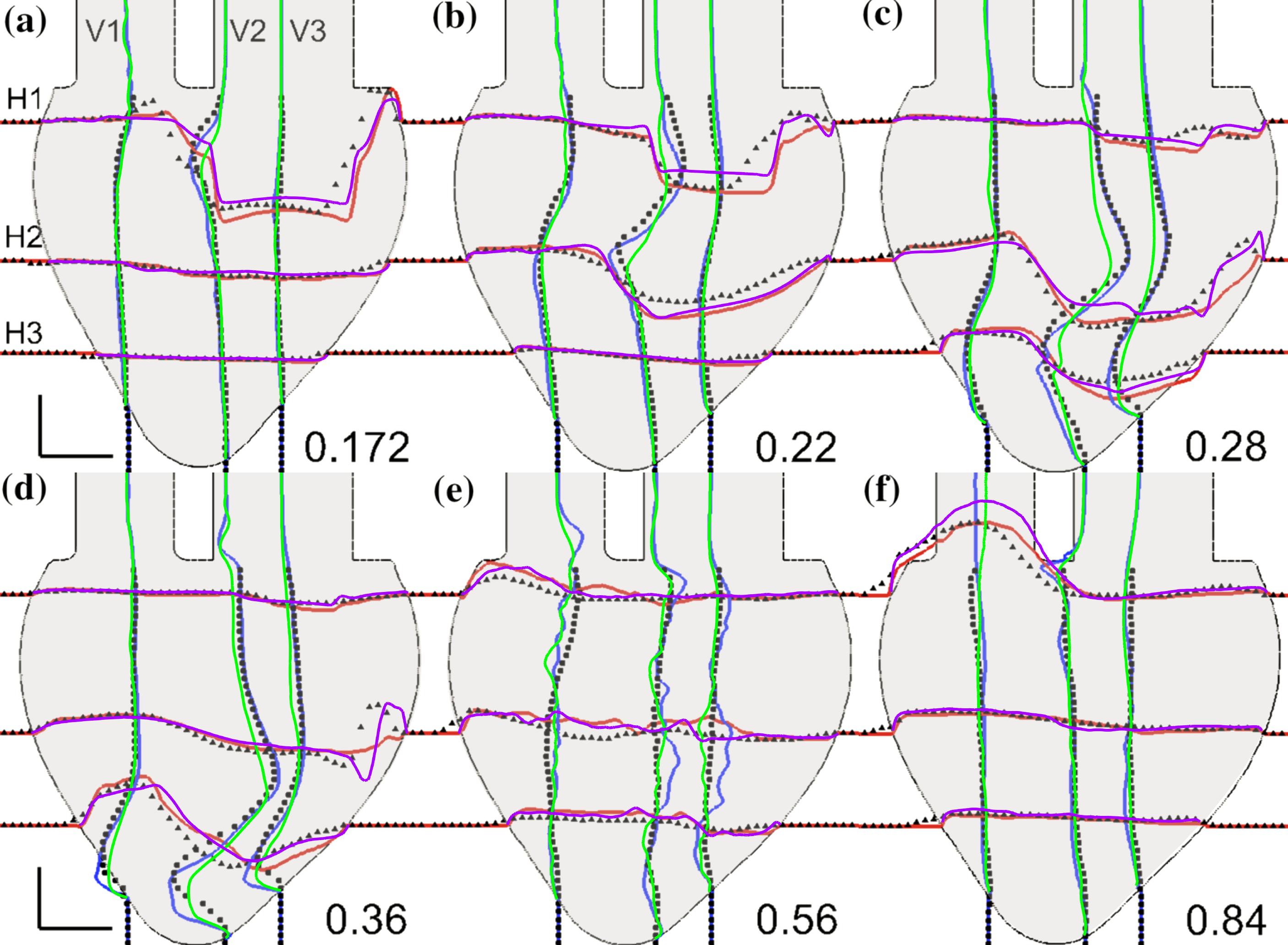} 
    \caption{
    Comparison of the phase-averaged horizontal and vertical components $v$ and $w$ scaled by their magnitude, between our computational results and that of Vedula et al.
    In our results, $v$ and $w$ are represented by the green and purple lines, respectively, while in the previous computational results, $v$ and $w$ are represented by the blue and red lines. 
    Previous experimental results are presented as symbols. 
    }
    \label{fig:vedula-velocity}
\end{figure}

In Figure \ref{fig:vedula-velocity} we present lateral and vertical velocity profiles, scaled by their magnitude and phase-averaged over five cycles, and shown in comparison to previous numerical and experimental results.
The lateral velocity components are plotted along three vertical lines crossing the LV lumen ($V1$, $V2$, $V3$), while the vertical velocity components are plotted along three horizontal lines ($H1$, $H2$, $H3$).
Our numerical results are plotted using purple and green solid lines, previous numerical results are plotted using red and blue solid lines, and previous experimental data is shown as symbols.
A qualitative comparison between our numerical and the previous experimental results indicate that our simulation reproduces the key features of the velocity components. 
During early diastolic phase, there is a slight discrepancy for the lateral velocity component which captures the inflow into the ventricle lumen. 
The primary differences throughout the diastolic phase is mostly visible in the vertical components, particularly at $t^* = 0.28$ and $t^* = 0.36$, where the differences are mostly apparent in the region near the lateral ventricle wall, similar to the numerical reference data.
Our results are generally less noisy compared to previous numerical results, in particular at $t^* = 0.56$, when our results deviate the least from the experimental results.

\begin{figure}
     \centering
     \begin{subfigure}[b]{0.46\textwidth}
         \centering
         \includegraphics[width=\textwidth]{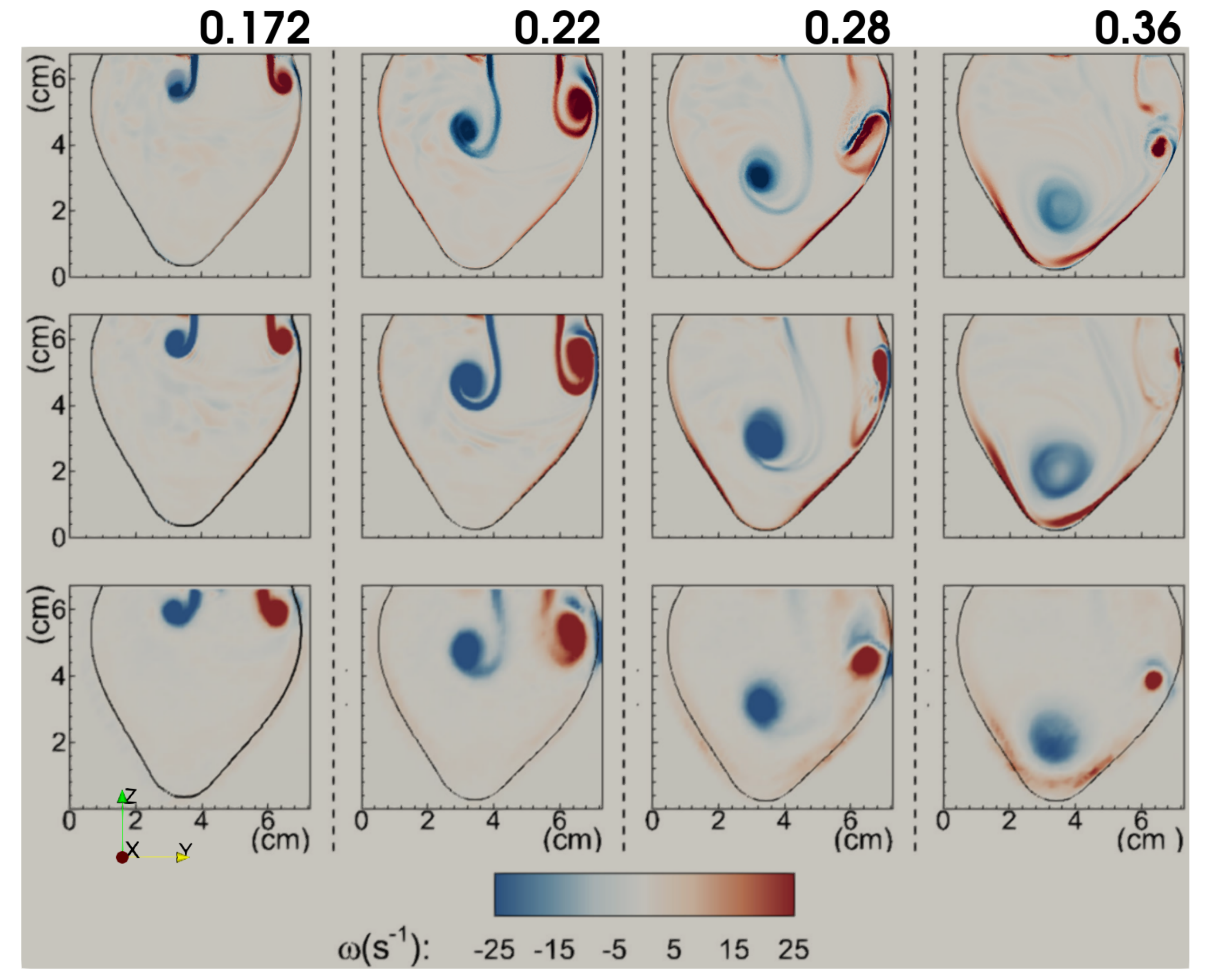}
         \caption{}
     \end{subfigure}
     \begin{subfigure}[b]{0.46\textwidth}
         \centering
         \includegraphics[width=\textwidth]{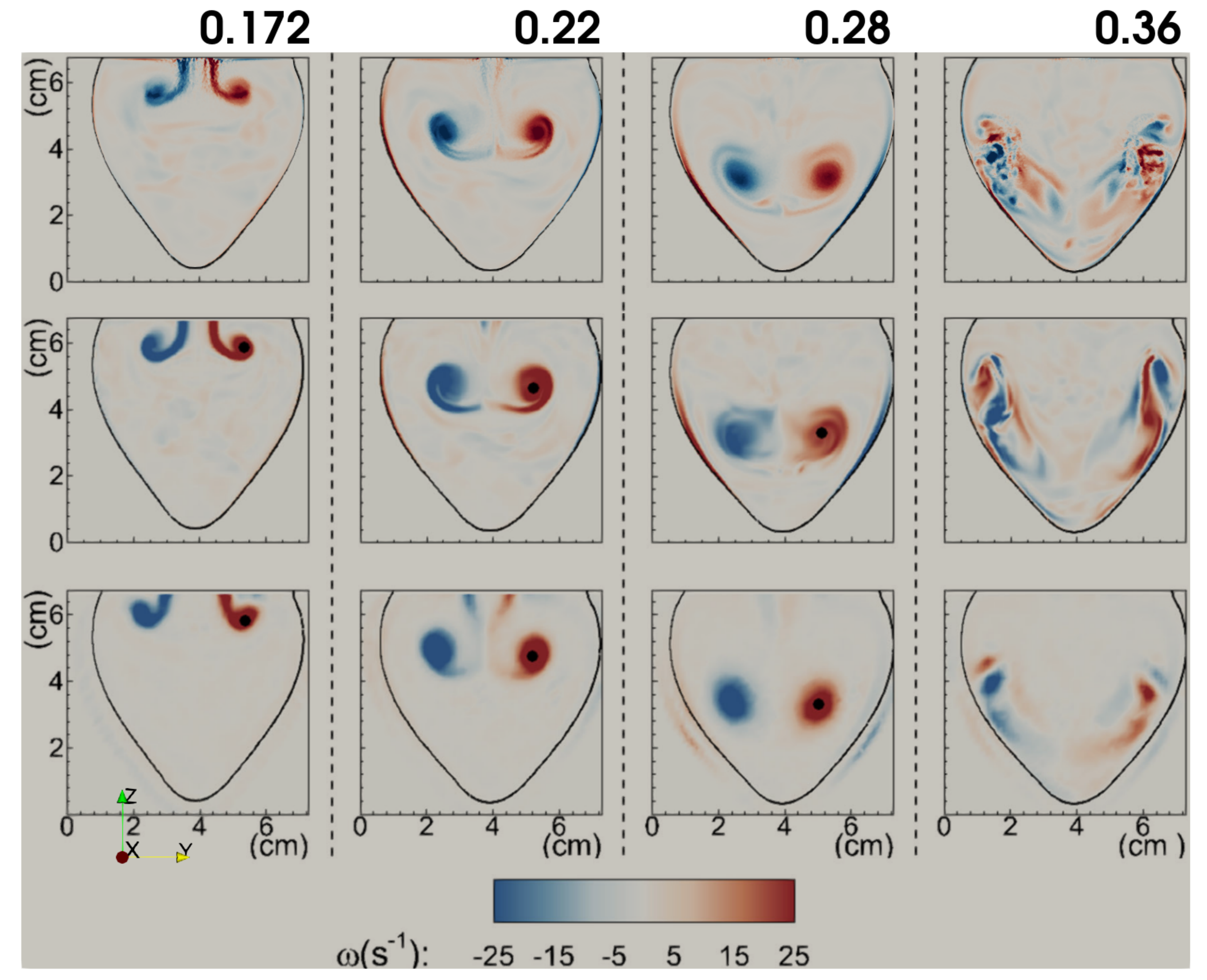}
         \caption{}
     \end{subfigure} 
      \caption{
      Comparison of the phase-averaged out-of-plane vorticity between our numerical results (top row) and that of Vedula et al., including numerical (middle row) and experimental (bottom row) results. 
      The results are presented in a frontal and sagittal slice of the ventricle in \textbf{(a)} and \textbf{(b)}, respectively. 
      The comparison is shown at four non-dimensional times throughout the cardiac cycle.}
    \label{fig:vedula-vorticity}
\end{figure}

In Figures \ref{fig:vedula-vorticity}\textbf{a} and \textbf{b}, we present the vorticity fields in the frontal and sagittal plane, respectively, sliced through the middle of the domain, and compared to previous numerical and experimental results. 
Focusing on Figure \ref{fig:vedula-vorticity}\textbf{a}, at $t^* = 0.172$, there is a qualitative similarity in the vorticity pattern, and agreement in vortical strength, although the left-most vortex in our simulation is slightly weaker in strength.
At the $t^* = 0.22$ both vortices have the characteristic shape as shown for the experimental results, with slight differences is strength. 
As the left (blue) vortex travels towards the LV apex in the subsequent frames, the right (red) vortex dissipates into a small but strong vortex near the ventricle wall, which is similar to the experimental results, but not captured by previous numerical results.

In Figure \ref{fig:vedula-vorticity}\textbf{b} the slices are captured in the sagittal plane, through the MV.
In terms of magnitude, phase, and location, the vorticity agrees reasonably well with the experimental results shown in the bottom row.
The simulations also capture the very slight tilt in the left (blue) vortex, which is observed in the experimental results at $t^* = 0.22$ and $t^* = 0.28$.
At $t^* = 0.36$ the vortex patterns appear to be dissipating, and harbouring a similar magnitude and location compared to experimental results. 

\section{Discussion}
\label{discussion}
We have shown that the OasisMove solver, where the N-S equations are expressed in the ALE formulation, produces numerically accurate solutions for a variety of fluid mechanics problems in moving domains. 
The solver has been tested and compared with existing results for the classic problem of flow around a oscillating cylinder, and for complex cardiovascular flow inside an idealized left ventricle, and was shown to produce fluid mechanically and physiologically plausible results. 

Code verification of the solver was performed by MMS analysis~\cite{roache2002code}, which rigorously assesses the solver's accuracy for a mathematically constructed problem in 2D.
As shown in the previous section, the theoretical convergence rates of $p=2$, $p=3$, and $q=2$ were achieved for the respective finite element polynomial orders, where $p$ measured the spatial rate, and $q$ measured the temporal rate. 
However, there are two observations that we will elaborate on.
First, in Figure \ref{fig:mms-ale-dx}\textbf{a} the mixed $\mathbb{P}_2/\mathbb{P}_1$ combination, often referred to as Taylor-Hood elements, resulted in a gradually steeper convergence rate for velocity as the mesh was refined, resulting in an improved convergence rate closer to $p=4$. 
Although we did not investigated this further, this behaviour could be related to the super-convergence for Taylor-Hood elements that has been previously observed using the standard incremental pressure correction scheme (IPCS) on regularly sized meshes aligned with the coordinate axis.~\cite{guillen2012superconvergence}
Nonetheless, the behaviour is not necessarily a deficiency of the solver considering the error was actually reduced.
Second, a dip in the pressure convergence rate plot was observed in Figure \ref{fig:mms-ale-dx}\textbf{b} for the $\mathbb{P}_1/\mathbb{P}_1$ combination, which results in a noticeable gap between the convergence rate lines.
Although we are not alarmed by such behaviour considering that the error continue to decrease rapidly, this gap may have been caused by round off errors, or the coarseness of the initial mesh resolutions.

Following verification, we selected flow past an oscillating cylinder to asses the solver's validity, a problem that is considered one of the most widely studied examples in fluid mechanics, with easily accessible and available reference data. 
Our results showed that the solver was able to reproduce lift and drag coefficients with less than 1\% error, capture the most dominant vortical patterns in the wake of the moving cylinder, and sufficiently compute the lift coefficient $C_L$ over multiple cycles for two distinct frequency ratios.  
There was however a slight discrepancy between the measured pressure coefficient $-C_{pb}$ and the reference data, as shown in Table \ref{tab:cylinder-moving}, resulting in an absolute error of 4.3\% for the finest mesh.
There may be multiple reasons for this deviation, one of which being related to a generally larger error in pressure compared to velocity, as already observed for the verification problem in Figure \ref{fig:mms-ale-dx}. 
Alternatively, the discrepancy could be due to measurement errors in the pressure values $p_0$ and $p_{180}$, which are pointwise quantities, in contrast to the integrated quantities $C_L$ and $C_D$.
Hence, the pointwise approximation could cumulatively cause pressure value offsets, resulting in the observed discrepancy in $-C_{pb}$.
It should be noted that our simulations were performed using $\mathbb{P}_1/\mathbb{P}_1$ elements, instead of the more accurate $\mathbb{P}_2/\mathbb{P}_1$ combination. 
Also, Blackburn and Henderson~\cite{blackburn1999study} used conforming quadrilateral elements, whereas we used linear triangular elements.
In addition, it has previously been shown that using the IPCS, in contrast to the consistent splitting scheme, introduces an artificial Neumann boundary condition, which in turn induces a numerical boundary layer that prevents the scheme from being fully second-order in pressure at the boundary~\cite{guermond2006overview}.
Therefore, for this particular problem, the pressure values on the boundary may be less accurate due to the presence of a no slip Dirichlet condition.   

The oscillating cylinder problem gave valuable insight into the solver's capability of correctly computing commonly reported flow characteristics and vortex dynamics at intermediate Reynolds numbers.
However, the problem was 2D laminar flow, and did not give insight into transitional or turbulent flow, which can be expected in physiological flows. 
Hence, we considered cardiovascular flow in an idealized 3D left ventricle, and compared our simulation results to both the numerical and experimental results of Vedula et al.~\cite{vedula2014computational}
A preliminary mesh convergence test was performed, showing a converging trend for all three velocity components.
The components showed qualitatively minor differences for the fine and very fine resolutions, and depicted physiologically plausible flows. 
Furthermore, for the total kinetic energy the mesh convergence test showed negligible qualitative difference in $KE^*$ between the fine and very fine resolution.
However, we observed a slightly larger gap between the medium and fine resolution in $KE^*$.
This gap may be due to under-resolved dissipation of the fluid at lower resolutions, considering that both the coarse and medium meshes displayed generally larger values of $KE^*$ than
was observed for the fine mesh. 
Following the mesh convergence test, a cycle convergence test was performed, which indicated little variation in the total $KE^*$ inside the LV cavity past the first cycle. 

The phase-averaged velocity and vorticity snapshots presented in Figures \ref{fig:vedula-velocity} and \ref{fig:vedula-vorticity}  proved to be qualitatively aligned with previous numerical and experimental results, where we sampled our data from a mesh of fine resolution (7M cells).
It should be noted that Vedula et al. used a computational grid of 128$\times$128$\times$256, and applied the immersed boundary method, where the computational model was immersed into a background Cartesian grid.
Hence, their computational grid and method is not directly comparable to our approach using the ALE formulation.
In practice, the immersed ventricle model would consist of far less fluid points that are part of the simulation.  
For instance, an ellipsoid immersed into a rectangular box would only fill half of its volume, resulting in 50\% non-fluid or "solid" cells, whereas the ALE based finite element strategy assures that all cells are part of the fluid solution.

It should also be noted that for both validation cases we deliberately did not consider temporal mesh refinement studies, but instead relied on satisfying the CFL condition, with CFL numbers less than or equal to 0.5 and 0.2 for the cylinder and ventricle problem, respectively.
This is comparable to the numerical simulations performed by Vedula et al.~\cite{vedula2014computational}, who used a maximum CFL number of about 0.3, while Blackburn and Henderson~\cite{blackburn1999study} did not specify the CFL number but instead reported that $0.0025 < \Delta t U_\infty / D < 0.004$ for the majority of their results.
Alternatively, we could have incorporated adaptive time steps depending on the spatial resolutions, also with the intention to satisfy a given CFL condition. 
However, through empirical evidence our results indicate that our strategy for selecting time step size assured adequate temporal accuracy. 

Although we have thoroughly investigated and demonstrated the performance of the solver for the selected test problems, this work has some limitations.    

First, for the 3D ventricle simulation we initially experienced numerical divergence at the open boundaries, when the fluid entered the domain with a purely wall driven flow.
We concluded that this divergence was caused by a downwind scheme effect, which has shown to be unconditionally unstable~\cite{sewell2005numerical}.
As a remedy, we prescribed a strongly imposed and mathematically determined Dirichlet velocity profile at the inlet based on the volume change, thus removing the downwind effect, and resulting in a stable solution, although it introduced one additional computation of the domain volume. 
Also, this downwind effect may have been present in all our simulations, although it was more pronounced for large Reynolds numbers, e.g. in the LV model where $Re = 3500$.

\begin{figure}
     \centering
     \begin{subfigure}[b]{0.45\textwidth}
         \centering
         \includegraphics[width=\textwidth]{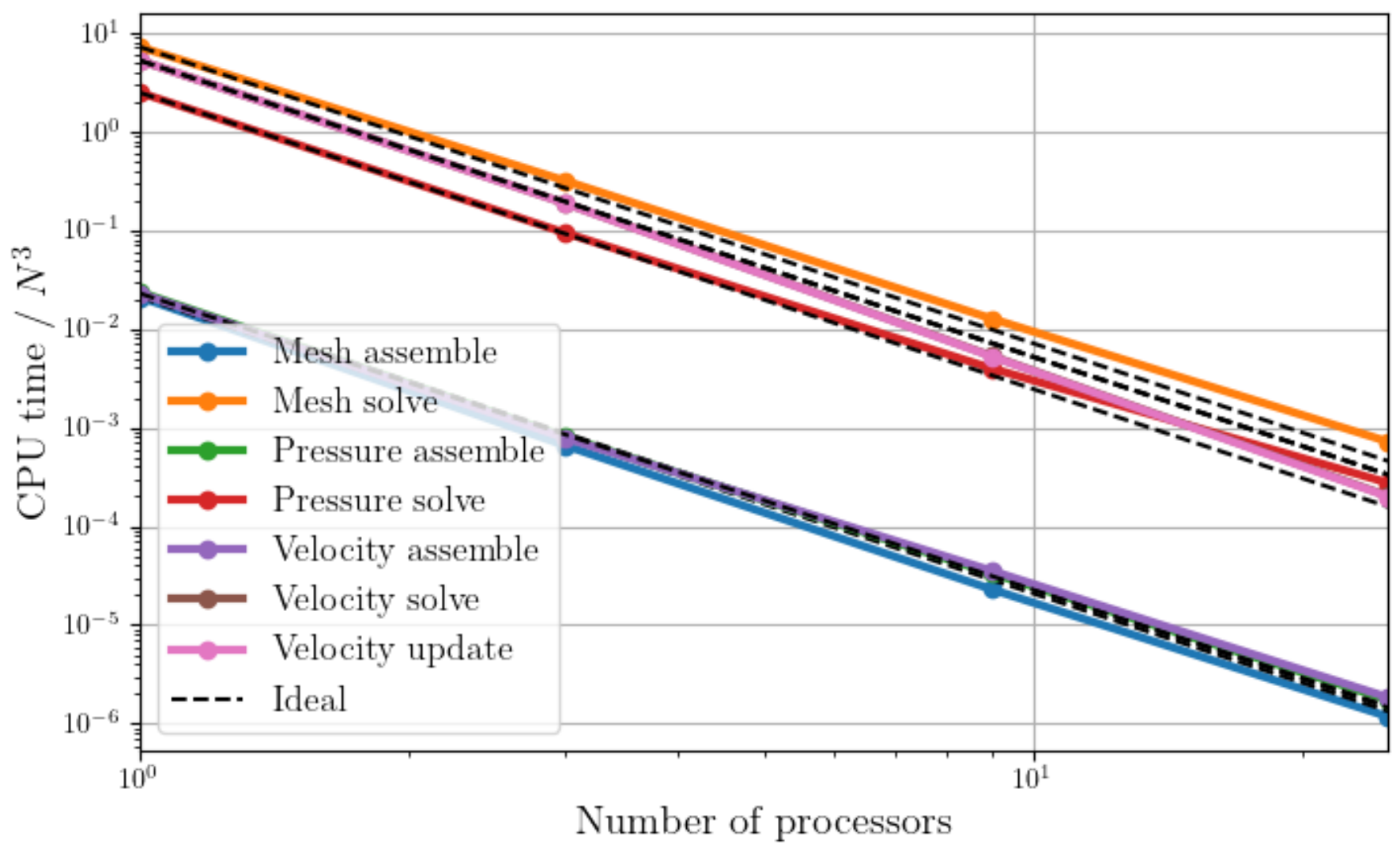}
         \caption{}
     \end{subfigure}
     \begin{subfigure}[b]{0.45\textwidth}
         \centering
         \includegraphics[width=\textwidth]{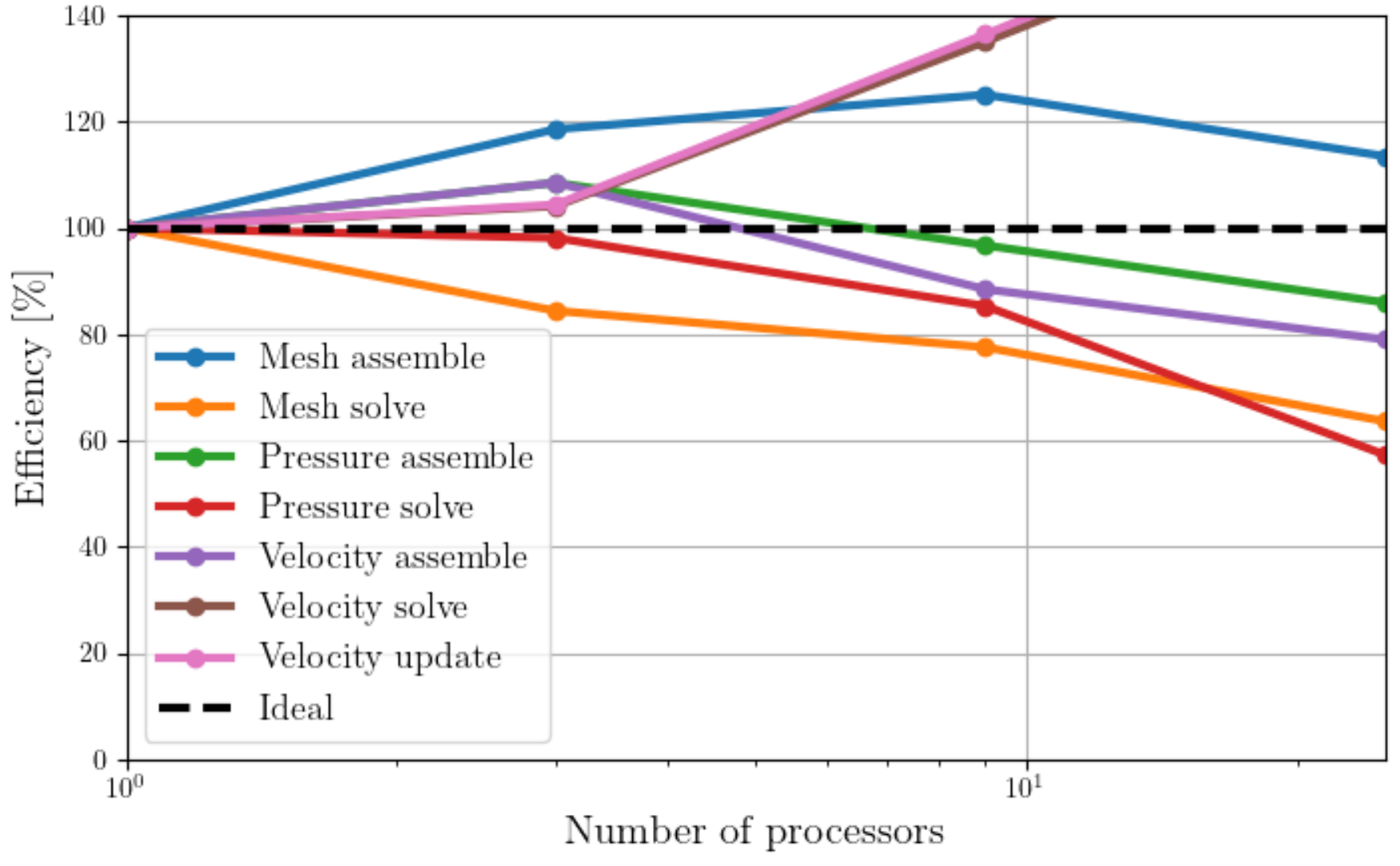}
         \caption{}
     \end{subfigure} 
  \caption{In \textbf{(a)}, we present the average CPU run time per time-step for the most essential parts of the OasisMove implementation scaled by the number of processors $N$, showing that the solver scales well weakly up to 25 CPUs on the Saga supercomputer. In \textbf{(b)}, we present the weak scaling efficiency as defined by \ref{eq:eff}, for the most essential parts of the solver.}
  \label{fig:vedula-weak}
\end{figure}

Second, we did not perform a comprehensive comparison against other NS-ALE or moving domain solvers, which could could give more insight into the relative performance and scalability of OasisMove. 
However, such a comparison is complicated by the fact that a number of solvers that would be natural to compare against are in-house, such as YALES2BIO or TUCAN~\cite{mendez2014unstructured,guerrero2017numerical}, others are commercial and require licenses, and available open-source solvers require extensive technical insight into their useage and underlying solution methods. We therefore
consider such a comprehensive study to be beyond the scope of this paper, but note that OasisMove's performance was shown to be on par with the original Oasis solver, even considering the additional computational time spent assembling and solving the mesh equation \eqref{eq:move}.

However, although a comprehensive study is left for future work, we have performed preliminary investigations into the software's scalability and entry-level high performance computing (HPC) capabilities.
We considered weak scaling of OasisMove using the 3D ventricle model considered in Section \ref{sec:3dval} as a representative real-world scenario. 
Using a time step of $\Delta t = 2\cdot 10^{-4}$, $\mathbb{P}_1/\mathbb{P}_1$ elements, and four models consisting of 800k, 2.5M, 7M, and 20M cells, we ran weak scaling on the authors' locally available HPC cluster for a total of 500 time steps per model, and computed the average run time per time step. 
For the mesh resolutions of interest, the weak scaling was run using 1, 3, 9, and 25 processors, respectively, resulting in approximately 800k cells per CPU. 
In Figure \ref{fig:vedula-weak}\textbf{a} we present the CPU run times scaled by the number of cores $N$ for the most essential parts of the software, showing that the most time consuming parts of the code include solving the pressure and mesh equation, and updating the velocity. 
In Figure \ref{fig:vedula-weak}\textbf{b} we present the weak scaling efficiency, defined as:
\begin{equation}
    \text{Efficiency} = \frac{t_1}{t_N}, \label{eq:eff}
\end{equation}
where $t_1$ is the amount of time to complete work with 1 processor, and $t_N$ is the amount of time to complete the same work using $N$ processors. 
For simplicity, we will focus on the mesh assembling and solving, which are the main extensions introduced in OasisMove.   
The weak scaling results show that assembling the mesh equations exhibit super-convergence, with an efficiency above $100\%$. 
However, it should be noted that the mesh assembly is considerably faster compared to solving the mesh equations, based on the CPU times in Figure \ref{fig:vedula-weak}\textbf{a}, and could be more sensitive to the cluster configuration and load.
When solving the mesh equations, the efficiency is above 60\% when running on 25 processors for the 20M cell mesh. 
However, note that these results are representative for this particular problem, and may vary depending on the problem complexity, and are also highly dependent on cluster design and load. 
Here, and for the majority of the presented problems, we performed simulations on Saga; a cluster with a total of 364 compute nodes, each with 40 2.0 GHz Intel Xeon-Gold processors.
For introductory problems, such as the Poisson and Stokes equation, the solver's backend has shown to perform with optimal efficiency~\cite{richardson2016high}, in contrast to the complex N-S equations, which may not be an optimal problem to assess efficiency.  
It should also be noted that the assembly and solving are strictly handled by low-level C++ implementations in FEniCS and PETSc~\cite{balay2019petsc}, which are minimally configurable with the exception for the choice of preconditioner and solver, and may be sensitive to cluster configuration.  
Regardless, the presented weak scaling results should give a good indication of the performance of OasisMove.
Furthermore, preliminary weak and strong scaling tests in comparison to default Oasis showed no noticeable differences in scalability and performance for a 2D flow problem on a rectangular domain.
Since the performance of the Oasis solver has been thoroughly demonstrated in previous works, these results indicate that the performance of OasisMove is comparable to the current state-of-the-art Navier-Stokes solvers.

\section{Conclusion}
\label{conclusion}
In conclusion, the OasisMove solver has shown high accuracy and performance agreement with theoretical convergence rates through verification, and has shown promising results in comparison to numerical and experimental data for the presented validation cases. 
With both prior and current in-depth knowledge on the solver's capabilities, the open-source CFD solver OasisMove is readily available for others to use and adapt. 
The range of problems implies that the solver is applicable to a wide range of problems within science and engineering, although we have specified its necessity and potential applicability to cardiovascular flows in bioengineering.  
Developing a set of benchmark problems for simulating cardiovascular fluid mechanics in moving cardiac domains is an important step in the process of verification and validation of cardiovascular CFD software, and may shed light on the challenges and limitations of such simulations.

\section*{Funding}
This work is a part of the ERACoSysMed PARIS project, and the SimCardioTest project, and has received funding from   
the European Union's Horizon 2020 research and innovation programme under Grant Agreements No 643271, and No 101016496, respectively.
The simulations were performed on the Saga cluster, with resources provided by UNINETT Sigma2 – the National Infrastructure for High Performance Computing and Data Storage in Norway, grant number nn9249k.

\section*{Acknowledgements}
The authors would like to thank Dr. Vijay Vedula, Prof. Rajat Mittal, and Prof. Giorgio Querzoli for their help to obtain computational models, and the permission to use both computational and experimental datasets from their previous research for the 3D validation problem presented in this paper~\cite{vedula2014computational}.

\section*{Conflict of interest}
The authors have no conflicts of interest.

\bibliography{ms}%
\end{document}